\documentclass[acmsmall,screen,nonacm]{acmart}

\usepackage{tikz}
\usetikzlibrary{shapes.geometric, arrows, positioning}
\AtBeginDocument{%
  }
\usepackage[most]{tcolorbox}
\usepackage{algorithmicx}
\usepackage{algpseudocode}
\definecolor{algoframe}{HTML}{666666}
\definecolor{algosection}{HTML}{999999}
\definecolor{algocomment}{HTML}{757575}
\algrenewcommand\algorithmiccomment[1]{\hfill{\color{algocomment}\textit{// #1}}}
\algrenewcommand\algorithmicif{\textbf{if}}
\algrenewcommand\algorithmicthen{\textbf{then}}
\algrenewcommand\algorithmicelse{\textbf{else}}
\algrenewcommand\algorithmicwhile{\textbf{while}}
\algrenewcommand\algorithmicreturn{\textbf{return}}
\tcbset{
  algosection/.style={
    enhanced,
    colback=white,
    colframe=algosection!40,
    arc=2pt,
    boxrule=0.4pt,
    borderline west={2pt}{0pt}{algosection},
    left=5pt, right=4pt, top=2pt, bottom=2pt,
    fonttitle=\small\sffamily\bfseries,
    coltitle=black,
    attach boxed title to top left={yshift=-1.5mm, xshift=3mm},
    boxed title style={colback=white, colframe=algosection!40, arc=1pt, boxrule=0.3pt},
  },
}

\setcopyright{none}
\copyrightyear{2026}
\acmYear{2026}
\acmDOI{}
\usepackage[binary-units]{siunitx}
\acmJournal{TACO}

\usepackage{multirow}
\usepackage{subfig}
\usepackage{graphicx}
\newif\ifanon
\anonfalse
\newcommand{\masked}[2]{\ifanon#2\else#1\fi}

\begin{document}

\title{Fast Forgiving Base64 Decoding with SIMD Instructions}
\author{Daniel Lemire}
\email{daniel.lemire@teluq.ca}
\orcid{0000-0003-3306-6922}
\affiliation{%
  \institution{Universit\'e du Qu\'ebec (TELUQ)}
  \city{Montreal}
  \state{Quebec}
  \country{Canada}
}

\renewcommand{\shortauthors}{Daniel Lemire}

\begin{abstract}
On the Internet, we use the Base64 format to represent binary data (e.g., images or cryptographic keys) as text. The ES2026 JavaScript standard, following the WHATWG \emph{forgiving-base64} rules, requires decoders to ignore ASCII whitespace while rejecting other non-alphabet characters.
Prior SIMD Base64 decoders assume whitespace-free input. We present a block-oriented decoding algorithm that removes whitespace inside the SIMD loop: clean 64-character blocks are decoded directly, while blocks with ignorable characters are compacted into a small stack buffer.
We provide implementations for x64 (SSSE3 through AVX-512~VBMI2), ARM~NEON, and LoongArch. Our C++ decoder reaches up to \SI{94}{\giga\byte\per\second} on recent AMD Zen~5 processors. In Node.js~25, integration of our decoder raises \texttt{Uint8Array.fromBase64} throughput by up to $6\times$ compared with Node.js~19. The implementation has been adopted by major browsers and JavaScript runtimes.
\end{abstract}



\settopmatter{printacmref=false}
\renewcommand\footnotetextcopyrightpermission[1]{}
\maketitle

\section{Introduction}
\label{sec:introduction}

Base64 is a binary-to-text encoding scheme that converts arbitrary binary data---images, files, any sequence of bytes---into a safe, printable ASCII string using a 64-character alphabet (A–Z, a–z, 0–9, +, /). We use it to send binary data by email~\cite{rfc2045} and inside JSON, HTML and XML documents.

Most general-purpose processors (x64, ARM, etc.) support single-instruction-multiple-data (SIMD) instructions~\cite{dubois2025pcod}.
Instead of operating on single words, SIMD instructions operate on several words at once. For example, you may add a vector of 16 words to another vector of 16 words using one instruction.
Mu{\l}a and Lemire~\cite{mula2020base64,mula2018faster} showed that---with SIMD instructions---we can encode and decode Base64 content at gigabytes per second: we may load 32~characters at once and produce 24~bytes using a handful of instructions.
Several libraries build on this work, among them \texttt{aklomp/base64}~\cite{aklomp2024base64} and Turbo-Base64~\cite{powturbo2023turbob64}; we survey them in Section~\ref{sec:related}.
They assume regular input: every byte is a Base64 character (or padding), with no whitespace or other special characters. Under that assumption the input can be loaded and decoded uniformly---each register-sized block of characters becomes a fixed number of output bytes, with no data-dependent control flow---which fits SIMD execution naturally.

That assumption does not hold on the web. Cryptographic keys and email attachments are stored in fixed-width lines separated by line feeds (`\textbackslash{}n'), and the WHATWG \emph{forgiving-base64} rules~\cite{whatwginfra} require a decoder to ignore ASCII whitespace while rejecting every other character outside the alphabet. Decoding is harder than encoding: any binary content is valid input to an encoder, but a decoder must validate the input and skip a sparse, irregularly placed subset of it.

Character by character, ignoring some characters is no concern: a branch per character suffices. But when characters are loaded in blocks we must detect and remove the ignorable ones from each newly loaded block. Pruning them in a separate pass before decoding is possible, but Mu{\l}a and Lemire report that such pruning can cost as much as the decoding itself~\cite{mula2018faster}. We find that a vectorized whitespace-removal pass in front of a fast clean-input decoder is usually slower than a decoder that does both at once. No SIMD Base64 decoder we know of removes ignorable characters inside its main loop; the \texttt{base64-simd}~\cite{nugine2022base64simd} library implements the same forgiving rules, but as a strip-then-decode pipeline.

\paragraph{The JavaScript Interface.}
These conventions are now part of a standard library API\@. JavaScript has a built-in type for arrays of 8-bit integers called \texttt{Uint8Array}. The JavaScript standard ECMAScript~2026 (ES2026) added \texttt{Uint8Array.fromBase64} and \texttt{Uint8Array.toBase64}, following the forgiving-base64 rules. They are supported by all major browsers (WebKit-based, Chromium-based, Firefox and more) and by major JavaScript runtimes (Node.js, Deno, Bun).
Encoding maps three bytes (24~bits) to four 6-bit values, each taken from the alphabet in Table~\ref{tab:base64table}. Because `\texttt{+}' and `\texttt{/}' have special meanings in URLs~\cite{nizipli2024parsing}, a URL variant uses `\texttt{-}' and `\texttt{\_}' instead. Decoding is the reverse. ASCII whitespace in the input is ignored~\cite{mdnuint8arrayfrombase64}. When the binary data is not divisible by three, we encode a single byte into two Base64 characters or two bytes into three, and we may add one or two padding characters (`\texttt{=}') so that the length is divisible by four.

\begin{table}\centering
\caption{Base64 mapping\label{tab:base64table} between 6-bit values and ASCII characters. The standard alphabet uses \texttt{+} and \texttt{/} for the last two characters while the URL variant uses \texttt{-} and \texttt{\_}.}%
\centering
\begin{tabular}{cc|cc|cc|cc|cc|cc}
\toprule
value & char & value & char & value & char & value & char & value & char & value & char \\
\midrule
0 & \texttt{A} & 1 & \texttt{B} & 2 & \texttt{C} & 3 & \texttt{D} & 4 & \texttt{E} & 5 & \texttt{F} \\
6 & \texttt{G} & 7 & \texttt{H} & 8 & \texttt{I} & 9 & \texttt{J} & 10 & \texttt{K} & 11 & \texttt{L} \\
12 & \texttt{M} & 13 & \texttt{N} & 14 & \texttt{O} & 15 & \texttt{P} & 16 & \texttt{Q} & 17 & \texttt{R} \\
18 & \texttt{S} & 19 & \texttt{T} & 20 & \texttt{U} & 21 & \texttt{V} & 22 & \texttt{W} & 23 & \texttt{X} \\
24 & \texttt{Y} & 25 & \texttt{Z} & 26 & \texttt{a} & 27 & \texttt{b} & 28 & \texttt{c} & 29 & \texttt{d} \\
30 & \texttt{e} & 31 & \texttt{f} & 32 & \texttt{g} & 33 & \texttt{h} & 34 & \texttt{i} & 35 & \texttt{j} \\
36 & \texttt{k} & 37 & \texttt{l} & 38 & \texttt{m} & 39 & \texttt{n} & 40 & \texttt{o} & 41 & \texttt{p} \\
42 & \texttt{q} & 43 & \texttt{r} & 44 & \texttt{s} & 45 & \texttt{t} & 46 & \texttt{u} & 47 & \texttt{v} \\
48 & \texttt{w} & 49 & \texttt{x} & 50 & \texttt{y} & 51 & \texttt{z} & 52 & \texttt{0} & 53 & \texttt{1} \\
54 & \texttt{2} & 55 & \texttt{3} & 56 & \texttt{4} & 57 & \texttt{5} & 58 & \texttt{6} & 59 & \texttt{7} \\
60 & \texttt{8} & 61 & \texttt{9} & 62 & \texttt{+} or \texttt{-} & 63 & \texttt{/} or \texttt{\_} \\
\bottomrule
\end{tabular}
\end{table}

\paragraph{Design Space.}
We might assume there are no ignorable characters and fall back on a slower approach once one is encountered. Unfortunately, a single whitespace character early in the input would then ruin the whole decode. A more nuanced approach falls back only locally, on the blocks where whitespace occurs. That is closer to workable, but two difficulties remain:
\begin{itemize}
    \item It scales poorly to wide registers. Recent processors have 64-byte SIMD registers, and we would like to work in units of 64~characters. Yet with line endings every 72~characters, most 64-character blocks contain whitespace, so the ``local'' fallback is the common case.
    \item Because four characters decode to three bytes, we cannot start decoding at an arbitrary offset. After removing whitespace, a 64-character block no longer holds a whole number of decodable groups, and completing the last one requires characters from the following block.
\end{itemize}

Thus, we developed the following strategy. We load the Base64 input in fixed-size blocks of 64~characters. We scan each block for errors and ignorable characters with SIMD instructions.
When a block contains no ignorable character---and nothing is pending from an earlier block---we decode it directly to the output. When ignorable characters are found,
we write the block's alphabet characters to a buffer on the stack, omitting the ignorable ones. Whenever the stack buffer approaches its capacity, we decode all of its complete 64-character groups to the output and slide the remainder to the front.
The stack buffer absorbs the misalignment that whitespace removal introduces, while the decoding step itself always sees uniform 64-character groups.

Our contributions are:
(1)~a block-oriented Base64 decoding algorithm that enforces WHATWG/\allowbreak ES2026 forgiving-base64 semantics---ignoring ASCII whitespace while reporting invalid characters---and does the whitespace removal inside the decoding loop rather than as a separate pass;
(2)~compaction and decoding kernels for x64 (SSSE3, AVX2, AVX-512~VBMI2), ARM~NEON, and LoongArch~LSX/LASX, including runtime instruction-set dispatch and specialized handling of single-whitespace blocks, together with a discussion of how the design maps onto SVE2 and RVV (Section~\ref{sec:scalable});
(3) an empirical study across ISAs, datasets, and whitespace densities that compares against the fastest available SIMD Base64 libraries (Section~\ref{sec:competitors}), reports hardware performance counters, decomposes the decoding time (Section~\ref{sec:timingprofile}), and ablates the main design parameters; and
(4) an open-source implementation integrated into major browsers and JavaScript runtimes.







\section{Base64 Conventions}
\label{sec:conventions}



In 2016, the Web Hypertext Application Technology Working Group proposed the WHATWG Infra Standard~\cite{whatwginfra}, which requires decoders to ignore ASCII whitespace characters but to report an error for unexpected characters. 
The ES2026 JavaScript standard has adopted the same conventions
for the \texttt{Uint8Array.fromBase64} function.
The C\# standard library (.NET) supports a similar convention, ignoring ASCII  whitespace characters.
Base64 is also used for data URLs. For example, we can embed an image directly into an HTML page
as \texttt{data:image/png;base64,iVBORw0K...} where \texttt{iVBORw0K...} is a Base64 encoded image. Browsers follow the WHATWG fetch conventions~\cite{whatwg-fetch} for data URLs, which rely on WHATWG Infra for the Base64 format. Thus ASCII whitespace characters are ignored within the data URL\@. Yet, if any characters other than the padding character, ASCII whitespace or Base64 characters are found, then the data URL is invalid. In such a case, no image would be displayed in the browser, or the icon of a broken image might appear.

The Base64 format was specified by Josefsson~\cite{rfc4648} in RFC~4648, which requires implementations to reject encoded data containing characters outside the base alphabet, whitespace included, unless explicitly stated otherwise. Java's \texttt{Base64.getDecoder().decode} follows RFC~4648; Python's \texttt{b64decode} follows it when its \texttt{validate} parameter is set and otherwise ignores non-Base64 characters. PHP's \texttt{base64\_decode} has a \texttt{strict} parameter that rejects non-alphabet characters other than whitespace; even in strict mode it still ignores ASCII whitespace. The MIME format~\cite{rfc2045}, used for email attachments, specifies lines of at most 76~characters and requires decoders to ignore line breaks and whitespace, while other characters may trigger a warning or a rejection.


\section{Related Work}
\label{sec:related}

\paragraph{Fast Base64 Encoding and Decoding.}
Mu{\l}a and Lemire~\cite{mula2018faster,mula2020base64} showed that SIMD instructions can encode and decode \emph{clean} Base64 input at gigabytes per second on x64 and AVX-512 processors. Their decoders assume that every input byte is a Base64 character (or padding). 

\paragraph{SIMD Base64 Libraries.}
Several libraries implement vectorized Base64; Table~\ref{tab:libraries} summarizes them and Section~\ref{sec:competitors} benchmarks them.
The \texttt{fastbase64}~\cite{lemire2018fastbase64} library is the reference implementation accompanying Mu{\l}a and Lemire~\cite{mula2018faster}, with SSE, AVX2 and AVX-512BW kernels on x64, NEON on 64-bit ARM and AltiVec on POWER.
The \texttt{aklomp/base64}~\cite{aklomp2024base64} library dispatches at runtime over SSSE3 through AVX-512 and NEON\@.
Turbo-Base64~\cite{powturbo2023turbob64} (also called \emph{base64-turbo}) targets the same instruction sets plus, recently, RISC-V. By default, Turbo-Base64 validates only partially; complete checking is a compile-time option, which we enable throughout our benchmarks. 
None of these three accepts ignorable whitespace: each rejects an embedded line feed as invalid. They can serve web conventions only behind a separate whitespace-removal pass.
The \texttt{base64-simd}~\cite{nugine2022base64simd} library is written in Rust with SSE4.1, AVX2, NEON and WebAssembly kernels. Its \texttt{forgiving\_decode} entry point implements the WHATWG rules we target. The implementation is a strip-then-decode pipeline: it locates the earliest ignorable character with a vectorized scan, removes the remaining whitespace with a scalar byte-by-byte pass into the destination buffer, then runs its vectorized clean-input decoder in place. 

\paragraph{SIMD Parsing and Validation.}
High-throughput parsers often classify bytes with SIMD comparisons, build bit masks, and compact or permute lanes to remove unwanted characters. Similar ideas appear in SIMD HTML scanning~\cite{lemire2025scanning} and in work on vectorizing interleaved or strided data~\cite{anderson2016interleaved}. Permutation instructions such as \texttt{pshufb} and \texttt{vpcompressb} are standard tools for within-register rearrangement~\cite{zhang2019poker}. Our work applies these techniques to Base64 decoding under validation rules that prior SIMD Base64 code did not support.

\begin{table}
  \caption{SIMD Base64 decoding libraries. The \emph{Forgiving} column indicates whether the library ignores ASCII whitespace as WHATWG Infra and ES2026 require; the \emph{fused} column indicates whether whitespace removal happens inside the decoding loop rather than as a separate pass. All entries are benchmarked in Section~\ref{sec:competitors}.}%
  \label{tab:libraries}
  \centering
  \footnotesize
  \setlength{\tabcolsep}{4pt}
  \begin{tabular}{llccl}
    \toprule
    Library & Instruction sets & Forgiving & Fused & Validation \\
    \midrule
    \texttt{fastbase64}~\cite{lemire2018fastbase64}     & SSE, AVX2, AVX-512BW, NEON, AltiVec              & no  & --- & complete \\
    \texttt{aklomp/base64}~\cite{aklomp2024base64}      & SSSE3--AVX-512, NEON                             & no  & --- & complete \\
    Turbo-Base64~\cite{powturbo2023turbob64}            & SSE, AVX, AVX2, AVX-512~VBMI, NEON, AltiVec, RVV & no  & --- & partial by default \\
    \texttt{base64-simd}~\cite{nugine2022base64simd}    & SSE4.1, AVX2, NEON, WASM                         & yes & no  & complete \\
    \textbf{this work}                                  & SSSE3, AVX2, AVX-512~VBMI2, NEON, LSX, LASX, AltiVec & yes & yes & complete \\
    \bottomrule
  \end{tabular}
\end{table}

\section{Our Algorithm}
\label{sec:algorithm}

We have multiple implementations of the same algorithm for different processor types.
The decoder is parameterized like the ES2026 \texttt{Uint8Array.fromBase64} API~\cite{mdnuint8arrayfrombase64}: callers choose an \emph{alphabet} and a \emph{last-chunk handling} mode.
Our C++ entry points expose the same choices as \texttt{base64\_default} versus \texttt{base64\_url} and as \texttt{loose}, \texttt{strict}, or \texttt{stop-before-partial}.
Section~\ref{sec:conformance} states the conformance semantics that the SIMD fast path and the scalar tail must satisfy; Fig.~\ref{fig:main} shows the block-oriented main loop once those semantics have been reduced to a byte stream of alphabet characters.
The algorithm requires a stack buffer of fixed size containing at least 128~bytes.

\subsection{Conformance Semantics}
\label{sec:conformance}

The decoder implements the WHATWG \emph{forgiving-base64} rules required by ES2026.
The \texttt{fromBase64} operation delegates to the ECMAScript \texttt{FromBase64} algorithm, which applies the WHATWG \emph{forgiving-base64} rules~\cite{whatwginfra}: ASCII whitespace is ignored, invalid characters are rejected, and the final incomplete 4-character chunk is handled according to \texttt{lastChunkHandling}. We implement the same observable behaviour, so that JavaScript engines can call our kernels without a semantic shim.

\paragraph{Ignorable Whitespace.}
Throughout decoding we treat the ASCII characters tab, line feed, form feed, carriage return, and space as ignorable~\cite{whatwginfra}. In the SIMD main loop (Fig.~\ref{fig:main}), each 64-character block is classified with two masks, non-alphabet versus whitespace; a character that is neither triggers an error carrying its offset. This is the fast-path realization of the specification's \texttt{SkipAsciiWhitespace} interleaved with chunk assembly.

\paragraph{Alphabet Selection.}
The \texttt{alphabet} option selects the standard alphabet (\texttt{+}/\texttt{/}) or the URL-safe one (\texttt{-}/\texttt{\_}) from Table~\ref{tab:base64table}. Under \texttt{base64url}, \texttt{+} and \texttt{/} are rejected while \texttt{-} and \texttt{\_} map to the same 6-bit values. All kernels share the same per-character lookup tables, compiled for the chosen alphabet.

\paragraph{Padding Discovery: An End-of-Input Preprocessing Step.}
The specification walks the input from beginning to end, skipping whitespace and assembling 4-character chunks; a \texttt{=} ends the current chunk and may introduce one or two padding bytes.
Padding can only occur in the closing chunk, so we address it initially by examining a short suffix of the input.
We call this the \emph{end-of-input preprocessing} step. It inspects only the last few characters, scanning them from the last character towards the beginning, and it never touches the rest of the input:
\begin{enumerate}
    \item Discard ignorable whitespace at the very end of the input.
    \item If the last remaining character is \texttt{=}, discard it, and then discard at most one further \texttt{=}, skipping any ignorable whitespace that separates the two.
    \item Record how many padding characters were discarded, and the position of the earliest \texttt{=} among them. This is not the position of the first \texttt{=} in the whole input: the step never scans past the short suffix it has examined.
\end{enumerate}
The suffix length is bounded by the amount of trailing whitespace plus two, so the cost is independent of the input length.

For example, \texttt{ABC=} and \texttt{ABC=\textbackslash{}n} are both well formed and decode to two bytes, and \texttt{AB==}, \texttt{AB==\textbackslash{}n} and \texttt{AB=\textbackslash{}n=} are all well formed and decode to one byte: ignorable whitespace may appear between or after the padding characters.
By contrast, \texttt{AB=A} and \texttt{A===} are ill formed: in the first the \texttt{=} does not sit in the closing chunk, and in the second three padding characters cannot complete a group of four.
The preprocessing step rejects both without a special case. Any \texttt{=} that is not part of the trailing padding run is neither an alphabet character nor whitespace, so the main loop's classification masks flag it as invalid. 
After this step the main loop sees an input whose length $l'$ excludes the trailing padding and the whitespace around it.

\paragraph{\texttt{lastChunkHandling}.}
After the main loop, at most one partial 4-character chunk remains (plus any bytes already compacted into the stack buffer).
The tail applies the three modes from ES2026:
\begin{itemize}
    \item \textbf{\texttt{loose}} (the default in \texttt{fromBase64}): a final chunk of two or three alphabet characters decodes to one or two output bytes; a lone trailing character is an error; if padding \texttt{=} characters were present, they must complete a full group of four symbols together with the preceding alphabet characters (whitespace ignored).
    \item \textbf{\texttt{strict}}: an unpadded final chunk of two or three characters is an error; if such a chunk is padded, any non-zero padding bits in the unused output byte(s) are an error (the specification's \texttt{throwOnExtraBits} check).
    \item \textbf{\texttt{stop-before-partial}}: an incomplete final chunk is left undecoded and, for length-reporting APIs, unconsumed---supporting \texttt{setFromBase64} and chunked decoding without guessing across buffer boundaries.
\end{itemize}

\paragraph{Length-Bounded Decoding.}
The \texttt{setFromBase64} method supplies a maximum output length. Our \texttt{base64\_to\_binary\_safe} entry point mirrors \texttt{FromBase64} with a \texttt{maxLength} argument: decoding stops once the destination is full and reports how many input code units were read, without writing past the bound.

\subsection{Block Preprocessing and the Main Loop}
\label{sec:blockloop}

After the end-of-input preprocessing above, the main loop of Fig.~\ref{fig:main} sees a plain stream of $l'$~characters that it processes in fixed blocks of sixty-four~characters. Each block goes through three stages.

\paragraph{Stage 1: Classification.}
After loading a block $b$ of 64~characters, we compute a 64-bit mask $m$ whose bit $i$ is set when $b[i]$ is not a character of the selected alphabet, using SIMD comparisons or table lookups (Section~\ref{sec:implementations}); on every architecture the same instructions that build $m$ also translate the alphabet characters into their 6-bit values, so classification and translation cost one pass, not two.
When $m$ is zero---the common case---every character belongs to the alphabet and no further mask is needed, at the cost of one test of a general-purpose register.
Only when $m \ne 0$ do we compute a second 64-bit mask $w$ marking the ASCII whitespace characters. If $m \wedge \lnot w \ne 0$, some character is neither an alphabet character nor ignorable whitespace, and we report an error at the offset of the lowest set bit. Otherwise $m = w$: the block is valid and $\mathrm{popcount}(m)$ characters must be removed.

\paragraph{Stage 2: Compaction.}
When $m \ne 0$ we write the $64 - \mathrm{popcount}(m)$  values to the stack buffer $S$, in order and without gaps: a byte-level \emph{stream compaction} governed by a bit mask, for which instruction sets offer very different support.
On AVX-512~VBMI2, \texttt{vpcompressb} performs exactly this in one instruction.
Elsewhere we build it as a permutation: the mask indexes a precomputed table of shuffle control vectors, and a \emph{shuffle} or \emph{permute} instruction applies it. Because a byte shuffle over a $k$-byte register would need a table of $2^{k}$ entries, and $2^{64}$ is out of the question, we decompose the 64-lane compaction into independent 8-lane compactions driven by an 8-bit sub-mask (a $256$-entry table), stitched together by a second lookup that closes the gaps between consecutive groups---the standard construction for byte compaction on SSE-class hardware~\cite{mula2018faster,zhang2019poker}.

Multiple ignorable characters in one block need no special treatment: the mask $m$ describes them all at once and the compaction removes them in the same permutation; only the number of bytes the block contributes to $S$ changes. For performance reasons, we specialize one case. With MIME-style line lengths---76~characters, or 72 for many mail agents---a 64-character block almost never contains two line breaks, so the common non-zero mask has a single set bit. When $m$ is a power of two we build the shuffle indices arithmetically instead of loading them (Section~\ref{sec:x64impl}); Section~\ref{sec:ablation} measures the benefit.

\paragraph{Stage 3: Decoding.}
The stack buffer never stores more than one byte per character---in most kernels it stores the 6-bit values produced in Stage~1 rather than the original characters---so a full stack buffer of $N \times 64$~bytes holds exactly $N$ decodable groups.
Whenever the free capacity of $S$ drops below 64~bytes, we decode all of its complete 64-byte groups to the output, then slide the fewer than 64 remaining bytes to the front of $S$. Each 64-byte group yields 48 output bytes.
When the block is clean \emph{and} $S$ is empty, the block never enters $S$ at all: it is decoded directly from the SIMD registers to the output. This is the fast path.

\paragraph{Why 64 Characters, on Every Register Width.} A block must be a multiple of four characters, since four characters decode to three bytes. The classification mask must fit in one general-purpose register, so that testing it, checking whether it is a power of two, and extracting the offset of an invalid character are single instructions; with one mask bit per character this caps the block at 64~characters on a 64-bit machine. And 64~bytes is the width of the widest SIMD register we use (AVX-512), so a 64-character block is one register there and a small fixed number of registers elsewhere: four 16-byte registers on NEON, SSSE3 and LSX, two 32-byte registers on AVX2 and LASX\@.
Using 64 everywhere therefore takes the largest block the mask allows and keeps the stack buffer and its flush granularity the same on every architecture. A narrower block---16~characters on NEON, say---would increase the mask tests and capacity checks per input byte.

\paragraph{Stack Buffer Size.}
The buffer must hold at least 128~bytes so that a flush always leaves room for the next block. Beyond that minimum, its size $N \times 64$ trades two costs against each other. Each flush decodes $N-1$ groups and then performs one memory copy of fewer than 64~bytes, so the amortized bookkeeping cost per decoded group falls like $1/(N-1)$; on the other hand, a very large buffer wastes cache and stack space. Section~\ref{sec:ablation} measures this trade-off and shows that the returns become negligible around $N = 8$, i.e. 512~bytes. We use $N=6$ (384~bytes) in the AVX-512 and LASX kernels and $N=10$ (640~bytes) in the NEON and LSX kernels.

\paragraph{Clean Input.}
Because of the design, an input with no ignorable character never touches the stack buffer: every block goes through classification and then straight to the output. Relative to our own clean path, that classification test is the only extra work. 

At the end of the processing, fewer than sixty-four~characters remain in the input, along with the remainder of the stack buffer.
The tail decoder applies the \texttt{lastChunkHandling} rules of Section~\ref{sec:conformance} to this remainder together with the padding count discovered during preprocessing.
We implement the tail with scalar code, except on AVX-512 hardware where SIMD processing is practical.
Because we expect most Base64 inputs to contain much more than 64~bytes, this last step does not dominate overall runtime.

\begin{figure}[t]
\centering
\begin{tcolorbox}[
  enhanced,
  colback=white,
  colframe=algoframe!50,
  arc=3pt,
  boxrule=0.6pt,
  left=8pt, right=8pt, top=6pt, bottom=6pt,
  width=0.92\linewidth,
  title={\textbf{Main loop of the Base64 decoder}},
  fonttitle=\small,
  coltitle=black,
  attach boxed title to top left={yshift=-2.5mm, xshift=4mm},
  boxed title style={colback=white, colframe=algoframe!50, arc=2pt, boxrule=0.4pt},
]
\small
\begin{tcolorbox}[algosection, title={Setup}]
\begin{algorithmic}[1]
  \State \textbf{require} input string $s$ of length $l$, output buffer $\sigma$ with capacity $\lceil 3l/4\rceil$
  \State apply alphabet selection and \texttt{lastChunkHandling} (Section~\ref{sec:conformance})
  \State preprocess $s$ to effective length $l'$ by stripping trailing padding \texttt{=} and surrounding whitespace
  \State $p \gets 0$;\quad $S \gets$ empty stack buffer (capacity $\geq 128$~bytes)
  \Statex \textit{Notation: $p$ is the read position in $s$, $b$ is the current block of 64 characters, $m$ and $w$ are 64-bit masks with one bit per character of $b$, marking respectively the non-alphabet and the whitespace characters, $|S|$ is the number of bytes currently held in $S$, and $\sigma$ is the output buffer.}
\end{algorithmic}
\end{tcolorbox}

\vspace{0.35em}
\begin{tcolorbox}[algosection, title={Hot loop}]
\begin{algorithmic}[1]
  \While{$p + 64 \leq l'$}
    \State $b \gets s[p{:}p{+}64)$;\quad $p \gets p + 64$ \Comment{64 characters}
    \State $m \gets \Call{NonAlphabetMask}{b}$ \Comment{SIMD}
    \If{$m \neq 0$}
      \State \hspace{1.15em}$w \gets \Call{WhitespaceMask}{b}$
    \EndIf
    \If{$m \neq 0$} \Comment{ignorables or invalid characters present}
      \If{$(m \wedge \lnot w) \neq 0$}
        \State \hspace{1.15em}\Return at first non-whitespace, non-alphabet offset
      \EndIf
      \State \hspace{1.15em}compact alphabet bytes of $b$ into $S$ (one byte per character; omit whitespace)
    \ElsIf{$|S| = 0$} \Comment{clean block, fast path}
      \State \hspace{1.15em}decode $b$ directly to $\sigma$ (64 characters $\to$ 48 bytes)
    \Else
      \State \hspace{1.15em}append $b$ to $S$ (one byte per character)
    \EndIf
    \If{$\Call{RemainingCapacity}{S} < 64$}
      \State let $k \gets |S|$
      \State \hspace{1.15em}decode first $\lfloor k/64\rfloor \times 64$ bytes of $S$ to $\sigma$
      \State \hspace{1.15em}slide the $k \bmod 64$ remainder to the front of $S$
    \EndIf
  \EndWhile
\end{algorithmic}
\end{tcolorbox}

\vspace{0.35em}
\begin{tcolorbox}[algosection, title={After the loop}]
\begin{algorithmic}[1]
  \State scalar \emph{tail} decoder: remaining $<64$ alphabet characters, stack remainder, and recorded padding count, with \texttt{lastChunkHandling} (Section~\ref{sec:conformance})
\end{algorithmic}
\end{tcolorbox}
\end{tcolorbox}
\Description{Pseudocode for the main loop of our Base64 decoding algorithm: setup, a 64-character SIMD hot loop with fast-path and stack-buffer branches, periodic stack compaction, and a scalar tail phase.}
\caption{\label{fig:main}Main loop of our Base64 decoding algorithm.}
\end{figure}

\section{Efficient Implementations}
\label{sec:implementations}

Different processor architectures (ARM, x64, LoongArch) have different SIMD instructions.
We implemented our algorithm for several architectures in C++ using \emph{intrinsic} functions, which usually map directly to a SIMD instruction; \texttt{vandq\_u8}, for instance, computes a bitwise AND between two ARM registers.
For x64 we provide three \emph{kernels}: a 128-bit one for Intel Westmere (2010) and AMD Bulldozer (2011) or better, a 256-bit AVX2 one for Intel Haswell (2013) and AMD Excavator (2015) or better, and a 512-bit AVX-512~VBMI2 one for Intel Ice Lake (2019) and AMD Zen~4 (2022) or better.
LoongArch, an instruction set developed by Loongson Technology and introduced in 2021 as an alternative to x64 and ARM, supports SIMD through LSX (128-bit) and LASX (256-bit); we provide a kernel for each.
On both families we detect the available instructions at runtime and always select the widest kernel the processor supports. This manual, multi-kernel strategy is a portable solution; frameworks and compiler tools can automate instruction selection~\cite{hohenauer2009simd}, but the manual approach was enough here.
For 64-bit ARM and for POWER (PPC64) with the ALTIVEC extension, a single implementation is chosen at compile time, and a portable C++ \emph{fallback} kernel covers everything else, so that all systems are supported. We make all our code freely available online.\footnote{\masked{\url{https://github.com/simdutf/simdutf}}{Repository URL omitted for anonymous review.}}

Our implementations accept UTF-8 (including ASCII) and UTF-16 input. For UTF-16, where each ASCII character occupies two bytes, we first check that the input is ASCII---any non-ASCII character is an error---and \emph{pack} the characters down to one byte each.

\subsection{LoongArch, POWER Altivec and x64 Implementations}
\label{sec:x64impl}

Mu{\l}a and Lemire describe in detail Base64 decoding on common x64 processors~\cite{mula2018faster}, without ignorable characters. We refer the interested reader to their article for further details.
On this basis, we implemented our algorithm 
for LoongArch, pre-AVX-512 x64, and POWER Altivec processors. For simplicity, we omit POWER Altivec processors from our discussion---it resembles LoongArch and x64.

All x64 processors by Intel and AMD support SIMD instructions (SSE2), but we implement fast decoding only from SSSE3 (2006) onwards, because we depend on \texttt{pshufb} (packed shuffle byte): a parallel 16-entry table lookup across the 16~bytes of a 128-bit register, indexed by the low 4~bits of a control byte per lane, which outputs zero in lanes whose control byte has its high bit set. AVX2 (2013) provides \texttt{vpshufb}, which acts like two \texttt{pshufb} calls, one per 128-bit half of a 256-bit register.
On LoongArch, \texttt{vshuf} (128-bit) and \texttt{xvshuf} (256-bit) select values from two SIMD registers using a third as indices. That is more powerful than \texttt{pshufb}---as are the corresponding instructions in AMD XOP, AVX-512 and POWER Altivec---but in practice we accomplish the same permutations either way. These parallel lookup instructions also serve to reorder elements within a register.

Our implementation operates on blocks of 64~input characters at a time, yielding 48 output bytes. But it loads four separate 16-byte chunks into four SIMD registers when using 16-byte registers (Westmere and LSX) or two separate 32-byte chunks when using 32-byte registers (LASX and AVX2).

We then convert Base64 characters into their 6-bit values in parallel according to Table~\ref{tab:base64table}. 
The core mechanism relies on a computed lookup using arithmetic and comparison operations to map ASCII characters (`A'--`Z', `a'--`z', `0'--`9', `+', `/', and optionally `-'/`\_' for the URL variant) to values 0--63, while detecting invalid characters. When needed, the identification of the whitespace characters is done similarly, with SIMD instructions.

We then map the results of these comparisons to a 64-bit mask. The SSE2 instruction \texttt{pmovmskb} extracts the most significant bit of each byte of a 128-bit register into a 16-bit integer; AVX2's \texttt{vpmovmskb} does the same for 32~bytes into a 32-bit integer. On LoongArch, \texttt{vmskltz} and \texttt{xvmskltz} are similar but write their result to a SIMD register, so an extra step is needed to move it to a general-purpose register.

To remove ignorable characters, we split the computed mask into 8-bit pieces and look each up in a $2^{8}$-element table of shuffle control vectors---up to four lookups for a 32-byte register---then apply them with a permutation instruction such as \texttt{pshufb}, packing the Base64 characters consecutively to the front of the register~\cite{zhang2019poker}. Though this general approach is fast, we also test whether the mask contains exactly one set bit---that is, whether the block holds a single ignorable character. Because a 64-bit mask $m$ is a power of two exactly when $m \wedge (m-1) = 0$, this test is two instructions. When it succeeds we compute the shuffle indices arithmetically instead of loading them, avoiding both the table load and its latency.

The computation is a compare-and-subtract. Let $j$ be the position of the character to remove, $\iota = (0, 1, \ldots, k-1)$ the identity vector, and $\mathbf{j}$ a vector with $j$ in every lane. The lane-wise comparison $\iota \geq \mathbf{j}$ yields a vector $g$ that is all-ones---that is, $-1$ read as a signed byte---at positions $\geq j$ and zero elsewhere, so $\iota - g$ adds one to every index at or beyond $j$ and leaves the others alone. On a four-lane register with $j = 2$, the vectors $\iota = (0,1,2,3)$ and $g = (0,0,-1,-1)$ give $\iota - g = (0,1,3,4)$, a shuffle that keeps lanes 0 and 1 in place and pulls lane~3 into position~2---exactly the removal of lane~2. The last index ($4$) points past the register; that output lane is meaningless but we never read it, since we advance the buffer pointer by only $k-1$ bytes.

The decoded 6-bit values are stored on the stack buffer, or used immediately. Before writing them out to the output, we pack them into consecutive bits. Each 16-byte register produces 12~output bytes; each 32-byte register produces 24~output bytes. This involves rearranging the bit fields from the naturally grouped 6-bit layout to a byte-aligned form. Efficient packing uses multiply-add instructions on x64 systems (e.g., to combine pairs of 6-bit values into 12-bit intermediates). On LoongArch, we use successive shifts (\texttt{vsrli} instruction) and AND/OR operations.

\subsection{ARM~NEON Implementation}
\label{sec:armimpl}

Like LoongArch and SSSE3-capable x64 processors, ARM NEON has a permutation instruction: \texttt{tbl}.
Since ARM NEON has no instruction that turns four SIMD registers into a 64-bit mask, we build one: we AND each of the four vectors with the repeating byte pattern (\texttt{0x01}, \texttt{0x02}, \texttt{0x04}, \ldots, \texttt{0x80}) that assigns a distinct bit to each lane within a group of eight, then apply horizontal pairwise byte sums to fold adjacent bytes together, repeating four times to reduce the 64~input bytes to a single 64-bit integer.

The decoding step also differs. ARM~NEON has \emph{structured} loads and stores that transfer $k$~registers at once while interleaving their bytes: loading two registers sends every other byte to the first and the rest to the second. We exploit this. Decoding loads the 6-bit values into four 16-byte vectors with \texttt{vld4}, which de-interleaves them into four streams; using such loads is an instance of the broader problem of vectorizing strided data layouts~\cite{anderson2016interleaved}. Left and right shifts (\texttt{shl}, \texttt{ushr}) combined with bitwise OR (\texttt{orr}) then recombine bits from adjacent vectors, inverting the encoding of 3~bytes into 4~characters, and \texttt{vst3} writes the result out while re-interleaving it.

\subsection{AVX-512 Implementation}
Mu{\l}a and Lemire describe Base64 decoding on AVX-512 processors without ignorable characters~\cite{mula2020base64}; we build on their approach.
Our AVX-512 implementation specifically requires VBMI2 (Vector Byte Manipulation Instructions~2), which accelerates byte-level compression, expansion and permutation within vectors---operations that would otherwise take several instructions each. VBMI2 is present on AMD Zen~4 and Zen~5 and on Intel Ice~Lake, Rocket~Lake, Sapphire~Rapids and Emerald~Rapids. Older AVX-512 processors lack it, and they also suffered from frequency throttling when issuing 512-bit instructions, which newer designs do not.
Here a 64-character block fits in a single register, so the kernel needs fewer instructions than the others.

For processing the input characters and deriving the 6-bit values, the AVX-512 VBMI instruction set has a powerful instruction that is ideally suited: \texttt{vpermi2b}. This instruction uses the least significant 7~bits of each byte to look up a value among 128~bytes, stored in two 64-byte registers.

We detect invalid characters as a by-product of this same lookup. We populate the 128-entry table so that every Base64 character maps to its 6-bit value (in the range 0--63, hence with a clear most significant bit), whereas every entry corresponding to a non-Base64 ASCII byte is set to a sentinel value whose most significant bit is set (e.g., \texttt{0x80}). Because \texttt{vpermi2b} only consults the least significant 7~bits, a non-ASCII byte (value $\geq 128$) could otherwise alias to a valid table entry; we therefore compute the bitwise OR of the translated vector with the original input bytes, which forces the most significant bit whenever the input byte itself has its high bit set. We then gather the most significant bit of each of the 64~bytes into a 64-bit mask using \texttt{vpmovb2m}. Any set bit in this mask marks a character that is not a Base64 character. When the mask is non-zero, we compute a second mask identifying the ASCII whitespace characters---using a \texttt{vpshufb}-based table comparison against the whitespace code points---and we take the exclusive OR of the two masks. A non-zero result indicates a character that is neither a Base64 character nor a whitespace character, and we report an error.

For removing ignorable characters, we use the AVX-512 VBMI2 instruction \texttt{vpcompressb}, which does exactly the work needed without the need for a lookup table.

\subsection{Extending the Design to Scalable Vector Extensions}
\label{sec:scalable}

The instruction sets above all have a fixed vector length known at compile time. Arm's SVE/SVE2~\cite{stephens2017arm,armsve2ref} and the RISC-V vector extension (RVV)~\cite{riscvvector} instead leave the vector length~$\mathrm{VL}$ unspecified, so a single binary runs on implementations from 128 to 2048~bits. We have not implemented kernels for either. (Although we have implemented Base64 \emph{encoding} for RVV.) The two extensions differ on the operation our algorithm depends on, so we set out how each stage of Section~\ref{sec:blockloop} would map onto them. Table~\ref{tab:scalable} summarizes the instructions.

\begin{table}
  \caption{What each stage of our algorithm needs: ``---'' marks the gaps that a scalable implementation would have to work around.}%
  \label{tab:scalable}
  \centering
  \small
  \begin{tabular}{llll}
    \toprule
    Requirement & AVX-512~VBMI2 & SVE2 & RVV~1.0 \\
    \midrule
    Classify bytes             & \texttt{vpermi2b}          & \texttt{TBL}/\texttt{MATCH}       & \texttt{vrgather.vv} \\
    Predicate from comparison  & \texttt{vpmovb2m}          & native (\texttt{svbool\_t})       & native (mask register) \\
    Predicate to scalar mask   & \texttt{kmovq}             & ---                                & \texttt{vmv.x.s} (on \texttt{vmsXX}) \\
    Count surviving bytes      & \texttt{popcnt}            & \texttt{CNTP}                     & \texttt{vcpop.m} \\
    Byte-granular compaction   & \texttt{vpcompressb}       & --- (32/64-bit \texttt{COMPACT} only) & \texttt{vcompress.vm} at \texttt{SEW=8} \\
    Bit packing 6\,$\to$\,8    & \texttt{vpmultishiftqb}    & \texttt{LD4B}/\texttt{ST3B} + shifts & \texttt{vlseg4e8}/\texttt{vsseg3e8} + shifts \\
    Tail handling              & \texttt{\_\_mmask64}       & \texttt{WHILELT}                  & \texttt{vsetvl} \\
    \bottomrule
  \end{tabular}
\end{table}

\paragraph{What Is Easier on Scalable Vectors.}
Three stages become simpler. Classification produces a predicate register directly---an SVE2 comparison yields an \texttt{svbool\_t}, an RVV comparison a mask register---so neither needs the movemask step that costs us several instructions on ARM~NEON (Section~\ref{sec:armimpl}) and an extra move on LoongArch; counting the survivors is likewise one instruction (\texttt{CNTP}, \texttt{vcpop.m}).
The partial final block disappears: our fixed-length kernels stop 64~characters before the end and hand the remainder to a scalar tail, whereas \texttt{WHILELT} on SVE2 or a \texttt{vsetvl} of the residual length on RVV lets the same vector code process the last, shorter block under a predicate. That removes a body of scalar code.

\paragraph{An Obstacle: Byte-Granular Compaction on SVE2.} In SVE2 the \texttt{COMPACT} instruction is defined only for 32-bit and 64-bit elements, with no byte-granular form. A byte compaction on SVE2 must therefore be synthesized, at a cost. Reusing the 256-entry shuffle tables of Section~\ref{sec:x64impl} requires materializing the predicate as an integer bit mask, and SVE has no movemask. SVE2.2 adds \texttt{COMPACT} support for 8-bit and 16-bit elements but no hardware supports SVE2.2 at this time to our knowledge.

\paragraph{Vector-Length Agnosticism and the Block Size.}
A second, more structural obstacle is that our 64-character block is pinned by the requirement that the classification mask fit in one 64-bit general-purpose register (Section~\ref{sec:blockloop}). Scalable code is free of that constraint but cannot exploit it either: at $\mathrm{VL} = 2048$~bits a block would naturally be 256~characters. A vector-length-agnostic kernel would therefore either keep 64-character blocks and leave lanes idle on wide implementations, or generalize the stack buffer so that its flush granularity is $\mathrm{VL}$ rather than 64~bytes.

Thus we leave RVV and SVE2 implementations to future work.


\section{Experiments}
\label{sec:experiments}

We evaluate our decoder on the seven systems of Table~\ref{tab:test-cpus} and the datasets of Table~\ref{tab:test-data}. We compare against the scalar decoders that browsers used to run, and against the fastest SIMD Base64 libraries we can find. We pay particular attention to inputs that contain ignorable characters, and we use hardware performance counters to explain the observed throughput differences across the x64, ARM, and LoongArch platforms.

We focus solely on decoding. In the JavaScript standard, the
encoding functions produce a Base64 string without ignorable
characters---a problem that received prior attention~\cite{mula2018faster,mula2020base64}.

\subsection{Software}

We implemented our algorithm  in C++ as part of \masked{the \texttt{simdutf} library}{a widely used open-source Unicode/SIMD library (name omitted for anonymous review)}. To help with reproducibility, we make both the library and all benchmarking code freely available under an open-source license.\footnote{\masked{\url{https://github.com/simdutf/simdutf/tree/master/benchmarks/base64}}{Repository URL omitted for anonymous review.}} The code is built using \textsc{CMake} with standard \textsc{Release} optimization flags. Performance is measured by repeatedly executing the decoding functions on each test input until the total runtime exceeds one second. We record both the best (peak) throughput and the average throughput across all runs, and we report the average, except in the timing profile (Section~\ref{sec:timingprofile}) and the scalability study (Section~\ref{sec:scalability}), which keep the fastest run. The median best-versus-average gap is about 1\,\% on every dataset; the largest outliers are competitor runs that put a scalar whitespace-removal pass in front of a SIMD decoder on the small logos. All measurements are performed in a single-threaded setting with no disk or network I/O.

We compare against two classes of alternative.
The first is the scalar decoders that our work replaces in production:
\begin{itemize}
    \item the {OpenSSL} Base64 decoder (version 3.3.x),\footnote{\url{https://www.openssl.org}}
    \item the conventional (pre-SIMD) Base64 decoder from Node.js~19.9.
\end{itemize}
The legacy Node.js Base64 decoder is not naive: it relies on tables to quickly convert and validate inputs. However, it was never used for the \texttt{Uint8Array.fromBase64} functions.
They are the practical baseline---they are what the web ran on before---but they are scalar, so a large speed-up over them is expected. We keep that comparison brief (Section~\ref{sec:cppresults}) and put the weight of the evaluation on the SIMD libraries.

The second class is the fast SIMD Base64 libraries of Table~\ref{tab:libraries}: \texttt{fastbase64}, \texttt{aklomp/base64} (version~0.5.2), Turbo-Base64, and \texttt{base64-simd} (version~0.8.0). Only the last implements forgiving semantics; the other three reject ignorable whitespace. A decoder for \texttt{Uint8Array.fromBase64} cannot assume the input is clean, so we never time those libraries on their own. We place each behind a removal pass, on every dataset including the Google logo, in two configurations:
\begin{itemize}
    \item behind a \emph{scalar} whitespace-removal pass, which copies the input to a scratch buffer while skipping ignorable characters with one branch per byte;
    \item behind a \emph{vectorized} whitespace-removal pass, which compacts the input using the same technique and the same tables as our own decoder---\texttt{vpcompressb} on AVX-512~VBMI2, the table-driven \texttt{tbl} permutation with the single-whitespace fast path on NEON, and the \texttt{vshuf} equivalent on LoongArch~LSX.
\end{itemize}
The 1.8~kB Bing logo is omitted from Tables~\ref{tab:competitors}--\ref{tab:competitors-aws-x64} because it duplicates googlelogo to within the run-to-run variation. We also report the removal pass on its own. For every dataset and configuration we check that the competitor produces byte-for-byte the same output as our decoder. All of them do.
Turbo-Base64 validates only partially in its default build: its unrolled loops check the first vector of each iteration and skip the rest. We compile it with complete validation (\texttt{B64CHECK}) so that every decoder in the comparison, on every machine, performs the same checking as ours. Further, Turbo-Base64 defers validation: it decodes regardless of errors (thus producing garbage) and does not report the position of the error.

\subsection{Correctness Validation}

We validate with more than one hundred unit tests per kernel, including TC39 \texttt{fromBase64} cases and regression tests for reported engine bugs, covering padding before and after \texttt{=}, partial final quanta under all three \texttt{lastChunkHandling} modes, both alphabets, embedded whitespace, invalid characters, and UTF-8 as well as UTF-16 input; every test is replayed on each SIMD kernel the host supports, including the portable fallback.
We also run \emph{differential} fuzzing: arbitrary byte strings are decoded under every combination of alphabet and last-chunk mode on all kernels, and any disagreement in output or error code aborts the run. Dedicated fuzz targets exercise the length-bounded safe API and encode--decode round trips, campaigns run on multiple ISAs, and the project participates in Google's OSS-Fuzz service. Production deployment in browsers and runtimes provides further real-world coverage.

\subsection{Hardware and Systems}
\label{sec:hardware}

Our experiments answer five questions:

\begin{enumerate}
  \item \emph{Does the design deliver across instruction sets?} (Section~\ref{sec:cppresults}) We use all seven systems of Table~\ref{tab:test-cpus}: three ARM, three x64 and one LoongArch, spanning client and server designs from Apple, Amazon, Intel, AMD and Loongson.
  \item \emph{Does it scale with SIMD register width?} (Section~\ref{sec:width}) This needs machines on which the \emph{same} algorithm runs at several widths. The three x64 systems support AVX-512~VBMI2 and so expose all three x64 kernels (128-bit SSSE3, 256-bit AVX2, 512-bit AVX-512~VBMI2) on one processor, and the Loongson-3A6000 exposes both LoongArch kernels (128-bit LSX, 256-bit LASX). The ARM systems have only the 128-bit NEON kernel.
  \item \emph{How does it compare with the fastest SIMD Base64 libraries?} (Section~\ref{sec:competitors}) The competing libraries have a different kernel per instruction set, so we need one representative of each: the Apple~M4~Max for ARM~NEON, the Intel Xeon~Gold~6548N for AVX-512~VBMI2, and the Loongson-3A6000 for LoongArch. We also ran the same harness on the four AWS metal instances (Graviton4, Graviton5, Granite Rapids, Zen~5).
  \item \emph{Where does the time go, and what do the design parameters buy?} (Sections~\ref{sec:timingprofile} and~\ref{sec:ablation}) These require rebuilding the decoder many times with modified constants, so we run them on the three machines we control directly---the Apple~M4~Max, the Intel Xeon~Gold~6548N and the Loongson-3A6000---which also give us a 128-bit, a 512-bit and a 256-bit kernel, enough to separate effects that depend on register width from those that do not.
  \item \emph{Does the benefit carry on to a JavaScript runtime?} (Sections~\ref{sec:nodejs} and~\ref{sec:browser}) This needs a production Node.js and browser installation: the Apple~M4~Max and the Intel Xeon~Gold~6548N.
\end{enumerate}

\noindent The systems themselves are specified in Table~\ref{tab:test-cpus}; we summarize here only the vector execution resources, since these bound the peak decoding throughput of a block-oriented SIMD decoder.

On ARM, our kernels use 128-bit NEON instructions, and all three ARM cores can retire four 128-bit vector operations per cycle. The Apple~M4 provides four 128-bit NEON pipelines. The Graviton4 uses the Arm Neoverse~V2 core (a server derivative of the Cortex-X3, Armv9.0-A) and the Graviton5 the newer Neoverse~V3 (Cortex-X4 generation, Armv9.2-A); both have four 128-bit SIMD/floating-point pipelines and, although both implement SVE2, both do so with a 128-bit vector length, identical to NEON~\cite{armneoversev2,armneoversev3}. The two Graviton instances (\texttt{c8g.metal-48xl} and \texttt{m9g.metal-48xl}) each expose 192~cores, the Graviton4 near \SI{2.8}{\GHz} and the Graviton5 near \SI{3.3}{\GHz}; since our benchmark is single-threaded, the salient differences are the clock and the per-core improvement of the V3 over the V2.

On x64, our widest kernel uses 512-bit AVX-512 registers and the VBMI and VBMI2 instructions (notably \texttt{vpermi2b} and \texttt{vpcompressb}); all three x64 processors support AVX-512~VBMI2 and all three have full-width 512-bit datapaths, so a 64-character block fits in one register. The Intel Xeon~Gold~6548N (Emerald~Rapids, Raptor~Cove cores) and the Intel Xeon~6975P-C (Granite~Rapids, Redwood~Cove cores) each expose two 512-bit FMA units; the AMD~EPYC~9R45 (Zen~5) implements AVX-512 over a \emph{native} 512-bit datapath---unlike Zen~4, which double-pumped a 256-bit unit---with four 512-bit vector pipelines and two 512-bit loads per cycle~\cite{yeezen5avx512}. 

On LoongArch, the Loongson-3A6000 is a quad-core, eight-thread LA664 design at up to \SI{2.5}{\GHz} on a 12~nm process, with 64~KB L1 instruction and data caches, a 256~KB private L2 and a shared 16~MB L3~\cite{loongson3a6000}. Each core is superscalar and out-of-order with four integer, four vector and four load/store units. The vector units implement LSX (128-bit) and LASX (256-bit); the 256-bit LASX width matches AVX2, while the Neoverse SVE2 implementations above are narrower at 128~bits. Although the Loongson-3A6000 has the lowest clock and core count of our systems, its per-clock performance is in the range of contemporary x64 cores.

We use performance counters to monitor the CPU frequencies and find them stable throughout our experiments; the AWS systems are bare-metal instances, so the counters reflect physical hardware rather than a virtualized guest. Frequency stability matters here because Section~\ref{sec:timingprofile} reports times rather than cycles. AVX-512 frequency throttling, which was substantial on Intel Skylake in 2017, has been absent or negligible since Ice Lake and never affected AMD parts; we could detect none on our hardware.

 \begin{table}
   \caption{Systems used for benchmarking}%
   \label{tab:test-cpus}
   \centering
   \begin{minipage}{\textwidth}
     \centering
\begin{tabular}{cccc}\toprule
  Processor & Max. Frequency & Microarchitecture & Compiler \\
  \midrule
  Apple M4 Max & \SI{4.5}{\GHz} & unnamed (aarch64, 2024) & Apple clang 17 \\
  AWS Graviton4 & \SI{2.8}{\GHz} & Neoverse V2 (aarch64, 2024) & GCC 16 \\
  AWS Graviton5 & \SI{3.3}{\GHz} & Neoverse V3 (aarch64, 2025) & GCC 16 \\
  Intel Xeon Gold 6548N & \SI{4.1}{\GHz} & Emerald Rapids (x64, 2023) & GCC 16 \\
  Intel Xeon 6975P-C & \SI{3.9}{\GHz} & Granite Rapids (x64, 2025) & GCC 16 \\
  AMD EPYC 9R45 & \SI{4.5}{\GHz} & Zen 5 (x64, 2024) & GCC 16 \\
  Loongson-3A6000 & \SI{2.5}{\GHz} & LA664 (loongarch64, 2023) & GCC 16 \\
  \bottomrule
\end{tabular}
   \end{minipage}
 \end{table}

 \begin{table}
   \caption{Datasets used for benchmarking}%
   \label{tab:test-data}
   \centering
   \begin{minipage}{\textwidth}
     \centering
     \begin{tabular}{lrrcc}\toprule
Name         &       Inputs  &  Base64 Chars  & Non-Base64 Chars  & Non-Base64 \% \\\midrule
DNS          &      \num{100000}   & \SI{34.4}{\mega\byte}  &     \SI{600}{\kilo\byte}   &         1.71 \\
email   &            15     &   \SI{1.95}{\mega\byte}    &     \SI{25.9}{\kilo\byte}        &    1.31   \\
googlelogo     &  1     &    \SI{3.14}{\kilo\byte}   &        0               &  0.00   \\
bing &           1   &    \SI{1.8}{\kilo\byte}    &        0   &              0.00    \\
       \bottomrule
     \end{tabular}
   \end{minipage}
 \end{table}

\subsection{Datasets}

We make all of our data available online.\footnote{\masked{\url{https://github.com/lemire/base64data}}{Repository URL omitted for anonymous review.}}
The \emph{DNS} dataset is the first \num{100000}~Base64 payloads of the \texttt{.se} (Sweden) zone file, each 350~characters in 7~sections separated by spaces; \emph{email} is fifteen Base64 files from the Enron corpus,\footnote{\url{https://www.cs.cmu.edu/~enron/}} in lines of 76~characters; and \emph{bing} and \emph{googlelogo} are the Base64-encoded logos of those sites, with no ignorable characters (Table~\ref{tab:test-data}).

\subsection{Absolute Throughput and the Scalar Baselines}
\label{sec:cppresults}

We measure decoding speed by repeatedly decoding the input for up to one second, at most a million times, and define speed from the input size: bytes of Base64 consumed per second. All functions decode into a pre-allocated buffer, so no allocation is timed. We report the average speed and use the relative difference between best and average as a measure of run-to-run variation. For our decoder the median gap is about 1\,\%; the largest we observed is 16\,\%, on the Apple~M4~Max logos.
Unless stated otherwise, each system runs its widest kernel: 512-bit AVX-512~VBMI2 on x64, 256-bit LASX on the Loongson-3A6000, 128-bit NEON on ARM.

Table~\ref{fig:performance} reports throughput on all seven systems, with the speed-up over the two scalar decoders we replace in production.
We reach up to \SI{18.4}{\giga\byte\per\second} on the Apple~M4~Max, \SI{7.6}{\giga\byte\per\second} on the Graviton4 and \SI{11.0}{\giga\byte\per\second} on the Graviton5; \SI{39.2}{\giga\byte\per\second} on the Intel~Xeon~Gold~6548N, \SI{48.9}{\giga\byte\per\second} on the Intel~Xeon~6975P-C and \SI{94.0}{\giga\byte\per\second} on the AMD~EPYC~9R45, the fastest of our systems; and \SI{6.0}{\giga\byte\per\second} on the Loongson-3A6000.
Against the scalar baselines this is 2.6--39.5 times the throughput of Node.js~19 and 8--102 times that of OpenSSL~3.3.x. A SIMD decoder is expected to beat scalar code. We report the ratios because they are what browsers and runtimes ran before the ES2026 API; Sections~\ref{sec:nodejs} and~\ref{sec:browser} show how much of the gap remains in JavaScript. Section~\ref{sec:competitors} compares against other SIMD decoders.

A more informative absolute reference is \texttt{memcpy}, which a Base64 decoder cannot beat since it reads four bytes for every three it writes. On the x64 servers we reach roughly a third to two thirds of the single-core \texttt{memcpy} bandwidth: on the AMD~EPYC~9R45, \SI{94.0}{\giga\byte\per\second} against \SI{145.3}{\giga\byte\per\second} on the Google logo and \SI{16.2}{\giga\byte\per\second} against \SI{30.7}{\giga\byte\per\second} on the larger DNS dataset.

\begin{table}
  \caption{Decoding speed of our decoder on all seven systems and four datasets, with the speed-up over the two scalar baselines it replaces ($\times$N over Node.js~19 and $\times$O over OpenSSL~3.3.x). Higher is better.}%
  \label{fig:performance}
  \centering
  \footnotesize
  \setlength{\tabcolsep}{3.5pt}
  \begin{minipage}{\textwidth}
    \centering

\begin{tabular}{lcccccccccccc}
\toprule
 & \multicolumn{3}{c}{bing} & \multicolumn{3}{c}{googlelogo} & \multicolumn{3}{c}{email} & \multicolumn{3}{c}{DNS} \\
\cmidrule(lr){2-4}\cmidrule(lr){5-7}\cmidrule(lr){8-10}\cmidrule(lr){11-13}
System & \si{\giga\byte\per\second} & $\times$N & $\times$O & \si{\giga\byte\per\second} & $\times$N & $\times$O & \si{\giga\byte\per\second} & $\times$N & $\times$O & \si{\giga\byte\per\second} & $\times$N & $\times$O \\
\midrule
Apple M4 Max & 18.1 & 3.9 & 25 & 18.4 & 4.0 & 25 & 14.7 & 3.7 & 21 & 9.4 & 2.7 & 14 \\
AWS Graviton4 & 7.3 & 3.5 & 15 & 7.6 & 3.6 & 16 & 6.2 & 3.5 & 13 & 4.2 & 2.6 & 9 \\
AWS Graviton5 & 10.3 & 3.6 & 13 & 11.0 & 3.8 & 13 & 8.4 & 3.7 & 11 & 6.2 & 3.0 & 8 \\
Intel Xeon Gold 6548N & 38.4 & 20.2 & 71 & 39.2 & 20.4 & 73 & 18.4 & 10.3 & 37 & 11.4 & 6.6 & 20 \\
Intel Xeon 6975P-C & 43.0 & 20.5 & 72 & 48.9 & 23.2 & 81 & 17.5 & 8.8 & 31 & 11.0 & 5.8 & 20 \\
AMD EPYC 9R45 & 72.1 & 30.5 & 85 & 94.0 & 39.5 & 102 & 29.7 & 13.0 & 41 & 16.2 & 7.3 & 23 \\
Loongson-3A6000 & 5.7 & 6.0 & 15 & 6.0 & 6.3 & 16 & 5.2 & 6.4 & 14 & 2.9 & 3.9 & 9 \\
\bottomrule
\end{tabular}
  \end{minipage}
\end{table}

 \begin{table}
   \caption{Performance-counter results: cycles per byte (c/b) and instructions per byte (i/b). Row labels abbreviate the benchmarked systems: M4 Max (Apple M4 Max), Grav.4 (AWS Graviton4), Grav.5 (AWS Graviton5), 6548N (Intel Xeon Gold 6548N), 6975P-C (Intel Xeon 6975P-C), 9R45 (AMD EPYC 9R45), and 3A6000 (Loongson-3A6000). OpenSSL is omitted; it executes 28--38 instructions per input byte on every system.}%
   \label{tab:perf-counters}
   \centering
   \begin{minipage}{\textwidth}
     \centering
     \small
     \setlength{\tabcolsep}{4pt}

\begin{tabular}{ll cc cc cc cc}
\toprule
 &  & \multicolumn{2}{c|}{bing} & \multicolumn{2}{c|}{googlelogo} & \multicolumn{2}{c|}{email} & \multicolumn{2}{c}{DNS} \\
 & & c/b & i/b & c/b & i/b & c/b & i/b & c/b & i/b \\
\midrule
\multirow{2}{*}{\small M4 Max} & node 19 & 0.88 & 7.60 & 0.87 & 7.58 & 0.99 & 8.75 & 1.13 & 9.40 \\
 & ours & 0.24 & 1.39 & 0.23 & 1.21 & 0.27 & 1.47 & 0.42 & 3.46 \\
\midrule
\multirow{2}{*}{\small Grav.4} & node 19 & 1.30 & 8.82 & 1.29 & 8.79 & 1.54 & 9.79 & 1.68 & 10.43 \\
 & ours & 0.37 & 1.31 & 0.36 & 1.20 & 0.44 & 1.68 & 0.65 & 3.19 \\
\midrule
\multirow{2}{*}{\small Grav.5} & node 19 & 1.15 & 8.82 & 1.13 & 8.79 & 1.43 & 9.79 & 1.58 & 10.43 \\
 & ours & 0.32 & 1.31 & 0.30 & 1.20 & 0.39 & 1.68 & 0.54 & 3.19 \\
\midrule
\multirow{2}{*}{\small 6548N} & node 19 & 1.83 & 11.56 & 1.82 & 11.54 & 1.94 & 12.40 & 2.03 & 12.95 \\
 & ours & 0.09 & 0.34 & 0.08 & 0.25 & 0.19 & 0.50 & 0.31 & 1.33 \\
\midrule
\multirow{2}{*}{\small 6975P-C} & node 19 & 1.85 & 11.57 & 1.85 & 11.54 & 1.96 & 12.46 & 2.05 & 13.07 \\
 & ours & 0.09 & 0.35 & 0.08 & 0.26 & 0.22 & 0.50 & 0.35 & 1.36 \\
\midrule
\multirow{2}{*}{\small 9R45} & node 19 & 1.91 & 11.57 & 1.89 & 11.54 & 1.97 & 12.46 & 2.04 & 13.07 \\
 & ours & 0.06 & 0.35 & 0.05 & 0.26 & 0.15 & 0.50 & 0.28 & 1.36 \\
\midrule
\multirow{2}{*}{\small 3A6000} & node 19 & 2.64 & 11.07 & 2.64 & 11.04 & 3.05 & 12.05 & 3.33 & 12.73 \\
 & ours & 0.44 & 2.07 & 0.42 & 1.95 & 0.48 & 2.45 & 0.85 & 4.13 \\
\bottomrule
\end{tabular}
   \end{minipage}
 \end{table}

Table~\ref{tab:perf-counters} shows that the scalar decoders are instruction-bound, executing 28--38 instructions per input byte (OpenSSL) and 7.6--13 (Node.js~19) on every platform, whereas our AVX-512 kernels use well under one instruction per byte on the clean datasets---about 0.3~i/b, 33--46 times fewer than Node.js---and our NEON and LASX kernels 1.5--4~i/b on email and DNS. 
\subsection{Scaling with SIMD Register Width}
\label{sec:width}

Because we select a kernel at runtime, we can run the \emph{same} algorithm at several SIMD register widths on one processor, isolating the benefit of wider vectors. Table~\ref{tab:width} reports the 128-bit, 256-bit and 512-bit kernels on the clean Google logo dataset, which fits in cache so the measurement is compute-bound. On x64, moving from 128-bit SSSE3 through 256-bit AVX2 to 512-bit AVX-512~VBMI2 improves throughput by $3.6$--$6.0\times$---from \SI{15.6}{\giga\byte\per\second} through \SI{29.6}{\giga\byte\per\second} to \SI{94.0}{\giga\byte\per\second} on the AMD~EPYC~9R45---and on LoongArch the 256-bit LASX kernel is about $1.7\times$ the 128-bit LSX one. Throughput therefore increases with register width, which is why we select the widest available kernel at runtime.

\begin{table}
  \caption{Decoding speed (\si{\giga\byte\per\second}) of the same algorithm at increasing SIMD register width, on the clean Google logo dataset. The widest available kernel is selected by default at runtime.}
  \label{tab:width}
  \centering
\begin{tabular}{lccc}\toprule
  Processor & 128-bit & 256-bit & 512-bit \\
  \midrule
  Intel Xeon Gold 6548N & 10.8 & 20.1 & 39.2 \\
  Intel Xeon 6975P-C & 12.2 & 22.8 & 48.9 \\
  AMD EPYC 9R45 & 15.6 & 29.6 & 94.0 \\
  Loongson-3A6000 & 3.5 & 6.0 & --- \\
  \bottomrule
\end{tabular}
\end{table}

\subsection{Comparison with Fast SIMD Base64 Libraries}
\label{sec:competitors}

We benchmark against the four SIMD libraries of Table~\ref{tab:libraries} on one system per instruction set: the Apple~M4~Max for ARM~NEON, the Intel Xeon~Gold~6548N for AVX-512~VBMI2, and the Loongson-3A6000 for LoongArch (Table~\ref{tab:competitors}). The same harness also ran on the four AWS metal instances (Tables~\ref{tab:competitors-aws-arm} and~\ref{tab:competitors-aws-x64}). These tables come from a different binary than Table~\ref{fig:performance}. On five of the seven systems the two agree to within a few percent---so the abstract's \SI{94}{\giga\byte\per\second} and the \SI{97.6}{\giga\byte\per\second} in Table~\ref{tab:competitors-aws-x64} are the same decoder on the same machine, measured two ways. The exceptions are Loongson DNS (18\,\%) and the Graviton4 (13--22\,\% across the three datasets). On x64, \texttt{fastbase64} is measured with its AVX2 kernel; Turbo-Base64 uses AVX-512~VBMI.

\paragraph{Semantics.}
We probed each library with inputs exercising the forgiving-base64 rules. The \texttt{fastbase64}, \texttt{aklomp/base64} and Turbo-Base64 libraries all reject an embedded line feed or space as invalid, so none can implement \texttt{Uint8Array.fromBase64} without an external whitespace-removal step. Compiled with complete validation, Turbo-Base64 rejects every invalid input we probe; its default partial-validation build instead accepts \texttt{SGV=bG8h} on our x64 machines, an input in which a padding character precedes the final chunk and which the specification requires to be rejected. The \texttt{base64-simd} library passes every probe, as does our decoder. On all four datasets, every competitor, in every configuration, produces byte-for-byte the same output as ours.

\paragraph{Clean Input.}
The Google logo contains no ignorable characters, but a forgiving decoder is not told that in advance, so every whitespace-unaware library still goes through a removal pass. Against the strongest such pipeline---Turbo-Base64 behind our vectorized strip---we are 1.7--3.5$\times$ faster on that dataset: \SI{6.6}{\giga\byte\per\second} against \SI{3.9}{\giga\byte\per\second} on the Graviton4, \SI{18.6}{\giga\byte\per\second} against \SI{7.7}{\giga\byte\per\second} on the Apple~M4~Max, \SI{40.1}{\giga\byte\per\second} against \SI{17.4}{\giga\byte\per\second} on the Xeon~Gold~6548N, \SI{49.1}{\giga\byte\per\second} against \SI{19.3}{\giga\byte\per\second} on Granite Rapids, and \SI{97.6}{\giga\byte\per\second} against \SI{28.0}{\giga\byte\per\second} on Zen~5. On LoongArch, where Turbo-Base64 has no vector kernel and falls back to scalar code, we are 2.3$\times$ on googlelogo (\SI{5.7}{\giga\byte\per\second} against \SI{2.5}{\giga\byte\per\second}). The extra pass over memory is the whole gap: our decoder still classifies the block, but it never leaves the direct path and never writes a stripped copy. 

\paragraph{Input with Whitespace.}
The email and DNS columns are the relevant ones for ES2026.
Against the strongest strip-then-decode baseline we can build---Turbo-Base64 preceded by a whitespace-removal pass using our own vectorized compaction---the picture depends on the machine. On ARM we lead every baseline, by 1.1--2.0$\times$: \SI{14.8}{\giga\byte\per\second} on email against \SI{7.6}{\giga\byte\per\second} on the Apple~M4~Max, \SI{8.3}{\giga\byte\per\second} against \SI{5.2}{\giga\byte\per\second} on the Graviton5, \SI{4.8}{\giga\byte\per\second} against \SI{3.3}{\giga\byte\per\second} on the Graviton4; the low end of the range is DNS on the Graviton4 (\SI{3.3}{\giga\byte\per\second} against \SI{3.0}{\giga\byte\per\second}). On the Intel Xeon~Gold~6548N we reach \SI{17.8}{\giga\byte\per\second} against \SI{12.1}{\giga\byte\per\second} on email. On the two AVX-512 AWS servers Turbo-Base64 behind our strip pass is close. On Granite Rapids DNS our average is \SI{10.74}{\giga\byte\per\second} against Turbo-Base64's \SI{11.13}{\giga\byte\per\second}, but the best rates are \SI{11.68}{\giga\byte\per\second} and \SI{11.19}{\giga\byte\per\second}: the 4\,\% gap on the mean is within run-to-run noise. On Zen~5 email we lead (\SI{27.3}{\giga\byte\per\second} against \SI{25.1}{\giga\byte\per\second}). With a scalar removal pass instead, every baseline is at most \SI{2.3}{\giga\byte\per\second} on the whitespace-containing datasets: the removal step has to be vectorized before the comparison is interesting.

Because the removal pass in these baselines is our own compaction code, the remaining difference is the extra pass over memory. The ``whitespace removal alone'' row prices that pass. On the Apple~M4~Max it runs at \SI{12.5}{\giga\byte\per\second} on email, which is slower than our complete decoder at \SI{14.8}{\giga\byte\per\second}: on that machine a strip-then-decode pipeline cannot beat us, no matter how fast its decoder. On the Intel machine the strip pass alone is faster than our decoder (\SI{20.2}{\giga\byte\per\second} against \SI{17.8}{\giga\byte\per\second}), so a faster decoder behind that pass could still win; the actual Turbo-Base64 pipeline does not (\SI{12.1}{\giga\byte\per\second}).

The \texttt{base64-simd} library is the only competitor that implements the same semantics. Its \texttt{forgiving\_decode} reaches \SI{2.0}{\giga\byte\per\second} on email and \SI{2.5}{\giga\byte\per\second} on DNS on the Apple~M4~Max, against our \SI{14.8}{\giga\byte\per\second} and \SI{9.5}{\giga\byte\per\second}, and on the Intel machine \SI{1.3}{\giga\byte\per\second} and \SI{1.6}{\giga\byte\per\second} against our \SI{17.8}{\giga\byte\per\second} and \SI{10.8}{\giga\byte\per\second}. Across all ARM and x64 machines the factor is 2.2--14$\times$, lowest on Graviton4 DNS\@. Its speed on whitespace-containing input is close to the scalar strip-then-decode rows of Table~\ref{tab:competitors}.

\begin{table*}
  \caption{Decoding speed (\si{\giga\byte\per\second}, higher is better) against the fast SIMD Base64 libraries, on one system per instruction set. Whitespace-unaware libraries are timed only behind a removal pass, including on googlelogo: a forgiving decoder cannot assume the input is clean. Turbo-Base64 ships no LoongArch kernel and falls back to scalar code on the Loongson-3A6000; \texttt{fastbase64} has no LoongArch kernel at all.}%
  \label{tab:competitors}
  \centering
  \footnotesize
  \setlength{\tabcolsep}{4pt}
  \begin{tabular}{lrrr@{\hspace{0.7em}}rrr@{\hspace{0.7em}}rrr}
\toprule
 & \multicolumn{3}{c}{\shortstack{Apple M4 Max\\[-2pt]{\scriptsize NEON}}} & \multicolumn{3}{c}{\shortstack{Intel Xeon Gold 6548N\\[-2pt]{\scriptsize AVX-512 VBMI2}}} & \multicolumn{3}{c}{\shortstack{Loongson-3A6000\\[-2pt]{\scriptsize LASX}}} \\
\cmidrule(lr){2-4}\cmidrule(lr){5-7}\cmidrule(lr){8-10}
Decoder & \emph{googlelogo} & \emph{email} & \emph{DNS} & \emph{googlelogo} & \emph{email} & \emph{DNS} & \emph{googlelogo} & \emph{email} & \emph{DNS} \\
\midrule
\textbf{this work (fused, forgiving)} & \textbf{18.6} & \textbf{14.8} & \textbf{9.5} & \textbf{40.1} & \textbf{17.8} & \textbf{10.8} & \textbf{5.7} & \textbf{5.2} & \textbf{2.4} \\
\addlinespace[2pt]
\multicolumn{10}{l}{\emph{whitespace-unaware libraries behind a scalar whitespace-removal pass}} \\
\quad\texttt{aklomp/base64} & 2.9 & 2.1 & 2.2 & 1.8 & 1.4 & 1.4 & 0.8 & 0.8 & 0.8 \\
\quad Turbo-Base64 & 2.9 & 2.1 & 2.2 & 2.0 & 1.4 & 1.6 & 0.9 & 0.9 & 0.8 \\
\quad\texttt{fastbase64} & 2.7 & 2.0 & 2.2 & 1.7 & 1.4 & 1.6 & --- & --- & --- \\
\addlinespace[2pt]
\multicolumn{10}{l}{\emph{same libraries behind our vectorized whitespace-removal pass}} \\
\quad\texttt{aklomp/base64} & 7.4 & 6.8 & 7.6 & 10.5 & 9.9 & 5.8 & 2.0 & 1.7 & 1.6 \\
\quad Turbo-Base64 & 7.7 & 7.6 & 7.9 & 17.4 & 12.1 & 9.5 & 2.5 & 2.1 & 1.8 \\
\quad\texttt{fastbase64} & 6.8 & 6.4 & 6.9 & 11.4 & 10.1 & 7.5 & --- & --- & --- \\
\midrule
\texttt{base64-simd} \texttt{forgiving\_decode} & 8.3 & 2.0 & 2.5 & 8.5 & 1.3 & 1.6 & --- & --- & --- \\
vectorized whitespace removal alone & 13.6 & 12.5 & 13.2 & 26.6 & 20.2 & 14.3 & 9.7 & 5.7 & 4.7 \\
\texttt{memcpy} & 114.0 & 83.7 & 67.9 & 137.9 & 25.0 & 31.6 & 51.0 & 21.7 & 15.3 \\
\bottomrule
\end{tabular}
\end{table*}

\begin{table*}
  \caption{Same comparison as Table~\ref{tab:competitors}, on the two AWS ARM servers. Units are \si{\giga\byte\per\second}.}%
  \label{tab:competitors-aws-arm}
  \centering
  \footnotesize
  \setlength{\tabcolsep}{4pt}
  \begin{tabular}{lrrr@{\hspace{0.7em}}rrr}
\toprule
 & \multicolumn{3}{c}{\shortstack{AWS Graviton4\\[-2pt]{\scriptsize NEON}}} & \multicolumn{3}{c}{\shortstack{AWS Graviton5\\[-2pt]{\scriptsize NEON}}} \\
\cmidrule(lr){2-4}\cmidrule(lr){5-7}
Decoder & \emph{googlelogo} & \emph{email} & \emph{DNS} & \emph{googlelogo} & \emph{email} & \emph{DNS} \\
\midrule
\textbf{this work (fused, forgiving)} & \textbf{6.6} & \textbf{4.8} & \textbf{3.3} & \textbf{11.1} & \textbf{8.3} & \textbf{6.1} \\
\addlinespace[2pt]
\multicolumn{7}{l}{\emph{whitespace-unaware libraries behind a scalar whitespace-removal pass}} \\
\quad\texttt{aklomp/base64} & 1.1 & 1.2 & 1.2 & 1.4 & 1.4 & 1.4 \\
\quad Turbo-Base64 & 1.2 & 1.1 & 1.2 & 1.4 & 1.5 & 1.5 \\
\quad\texttt{fastbase64} & 1.1 & 1.1 & 1.1 & 1.3 & 1.3 & 1.4 \\
\addlinespace[2pt]
\multicolumn{7}{l}{\emph{same libraries behind our vectorized whitespace-removal pass}} \\
\quad\texttt{aklomp/base64} & 3.8 & 3.3 & 2.8 & 4.9 & 4.8 & 4.3 \\
\quad Turbo-Base64 & 3.9 & 3.3 & 3.0 & 5.3 & 5.2 & 4.5 \\
\quad\texttt{fastbase64} & 3.0 & 2.7 & 2.5 & 3.9 & 3.8 & 3.7 \\
\midrule
\texttt{base64-simd} \texttt{forgiving\_decode} & 3.8 & 1.3 & 1.5 & 5.1 & 1.9 & 2.1 \\
vectorized whitespace removal alone & 7.7 & 6.1 & 5.5 & 9.8 & 9.4 & 8.6 \\
\texttt{memcpy} & 48.7 & 28.9 & 22.4 & 102.2 & 57.8 & 48.8 \\
\bottomrule
\end{tabular}

\end{table*}

\begin{table*}
  \caption{Same comparison as Table~\ref{tab:competitors}, on the two AWS x64 servers. Units are \si{\giga\byte\per\second}. On x64, \texttt{fastbase64} uses its AVX2 kernel.}%
  \label{tab:competitors-aws-x64}
  \centering
  \footnotesize
  \setlength{\tabcolsep}{4pt}
  \begin{tabular}{lrrr@{\hspace{0.7em}}rrr}
\toprule
 & \multicolumn{3}{c}{\shortstack{Intel Xeon 6975P-C\\[-2pt]{\scriptsize AVX-512 VBMI2}}} & \multicolumn{3}{c}{\shortstack{AMD EPYC 9R45\\[-2pt]{\scriptsize AVX-512 VBMI2}}} \\
\cmidrule(lr){2-4}\cmidrule(lr){5-7}
Decoder & \emph{googlelogo} & \emph{email} & \emph{DNS} & \emph{googlelogo} & \emph{email} & \emph{DNS} \\
\midrule
\textbf{this work (fused, forgiving)} & \textbf{49.1} & \textbf{18.3} & \textbf{10.7} & \textbf{97.6} & \textbf{27.3} & \textbf{15.2} \\
\addlinespace[2pt]
\multicolumn{7}{l}{\emph{whitespace-unaware libraries behind a scalar whitespace-removal pass}} \\
\quad\texttt{aklomp/base64} & 2.7 & 1.9 & 1.7 & 3.1 & 2.1 & 2.3 \\
\quad Turbo-Base64 & 3.1 & 2.0 & 1.9 & 3.4 & 2.2 & 2.3 \\
\quad\texttt{fastbase64} & 2.8 & 1.9 & 1.8 & 3.2 & 2.1 & 2.3 \\
\addlinespace[2pt]
\multicolumn{7}{l}{\emph{same libraries behind our vectorized whitespace-removal pass}} \\
\quad\texttt{aklomp/base64} & 11.2 & 10.5 & 7.1 & 16.1 & 17.2 & 8.7 \\
\quad Turbo-Base64 & 19.3 & 12.2 & 11.1 & 28.0 & 25.1 & 12.5 \\
\quad\texttt{fastbase64} & 12.3 & 10.4 & 8.3 & 18.1 & 17.6 & 10.8 \\
\midrule
\texttt{base64-simd} \texttt{forgiving\_decode} & 9.1 & 1.4 & 1.5 & 12.8 & 2.4 & 2.4 \\
vectorized whitespace removal alone & 28.0 & 20.0 & 17.3 & 37.9 & 37.6 & 20.0 \\
\texttt{memcpy} & 118.2 & 24.6 & 26.9 & 143.2 & 79.3 & 30.2 \\
\bottomrule
\end{tabular}

\end{table*}

\subsection{Timing Profile: Where Does the Decoding Time Go?}
\label{sec:timingprofile}

\begin{figure}
\centering
\subfloat[Decoding time per input byte against $f$, obtained by varying the line length; dashed lines are the fits of Eq.~\ref{eq:profile} over line lengths $L \geq 64$. The hollow markers at $f = 0$ are inputs with no line breaks.\label{fig:profile}]{\includegraphics[width=0.48\textwidth]{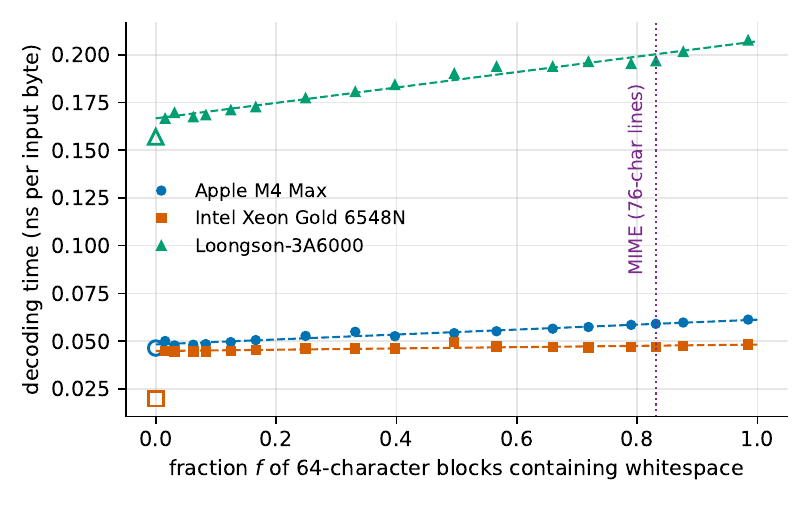}}
\hfill
\subfloat[Decoding speed against stack buffer size at MIME line length. \label{fig:bufsweep}]{\includegraphics[width=0.48\textwidth]{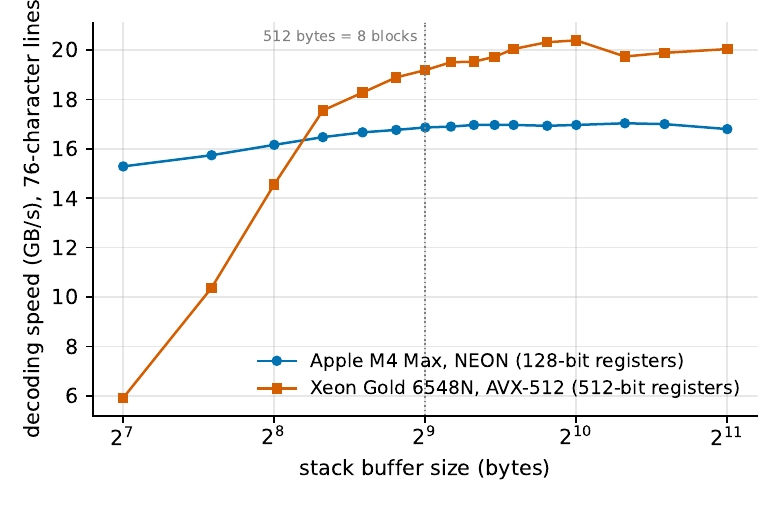}}
\Description{Left: decoding time per byte versus the fraction of blocks containing whitespace, with linear fits, for three systems. Right: decoding speed versus stack buffer size for a NEON and an AVX-512 kernel, both flattening near 512 bytes.}
\caption{Timing profile (a) and stack-buffer ablation (b).}
\label{fig:profilepair}
\end{figure}

The main loop takes one of two paths per 64-character block. Let $f$ be the fraction of blocks that contain at least one ignorable character. With a line break every $L$ characters, $f = \min(1, 64/(L+1))$. We encode a fixed \SI{256}{\kilo\byte} binary payload---small enough to stay in cache, so the measurement is compute-bound---at line lengths from 16 to 4096. We also add one input with no line breaks. We measure the decoding time per input byte, keeping the fastest of repeated runs (Fig.~\ref{fig:profile}, Table~\ref{tab:profile}); a throughput in \si{\giga\byte\per\second} is then $1/t$. For line lengths $L \geq 64$, where a block contains at most one line break and $0 < f < 1$, the measurements lie on a straight line,
\begin{equation}
  t(f) = t_{\mathrm{buf}} + f\,\Delta ,
  \label{eq:profile}
\end{equation}
to within the run-to-run variation. The variable $\Delta$ is the extra cost of a block that contains at least one ignorable character---the compaction of Stage~2. The variable $t_{\mathrm{buf}}$ is the intercept of that fit. Write $t_{\mathrm{fast}}$ for the cost of a clean block on the direct fast path (the hollow markers at $f = 0$). The direct fast path requires an empty stack buffer, and after the first ignorable character the buffer almost never holds a multiple of 64~bytes, so that path is unavailable for the rest of the input.

\paragraph{AVX-512.}
On the Intel Xeon~Gold~6548N we measure $\Delta = \SI{0.0034}{\nano\second\per\byte}$: small next to a clean block, since \texttt{vpcompressb} does the Stage-2 compaction in one instruction. A clean block costs $t_{\mathrm{fast}} = \SI{0.0199}{\nano\second\per\byte}$ on the fast path, i.e.\ \SI{50.3}{\giga\byte\per\second}. The intercept $t_{\mathrm{buf}} = \SI{0.0448}{\nano\second\per\byte}$ (\SI{22.3}{\giga\byte\per\second}) is about twice that, so at MIME line lengths almost all of the whitespace penalty is the loss of the direct-to-output path.
We confirmed this by disabling the fast path and rebuilding: on the same whitespace-free input, the throughput falls from \SI{50.3}{\giga\byte\per\second} to \SI{29.8}{\giga\byte\per\second}, while every $f > 0$ measurement is unchanged to within 2\,\%, as expected if that path was already unused whenever $f > 0$. (The two builds differ only in that path and ran alternately.)
Without the fast path, a whitespace-free block still goes through the stack buffer (\SI{29.8}{\giga\byte\per\second}), but it is faster than $1/t_{\mathrm{buf}}$ (\SI{22.3}{\giga\byte\per\second}): $t_{\mathrm{buf}}$ is fitted on input that contains whitespace, so each block contributes fewer than 64~bytes, the buffer flushes at irregular offsets, and the copy that slides the remainder to the front is no longer of length zero. The three rates, \SI{50.3}{}, \SI{29.8}{} and \SI{22.3}{\giga\byte\per\second}, therefore price the direct path, the stack-buffer round trip, and the misaligned flushes in turn.

\paragraph{NEON and LASX.}
On the Apple~M4~Max, $t_{\mathrm{fast}} = \SI{0.0464}{\nano\second\per\byte}$ and $t_{\mathrm{buf}} = \SI{0.0483}{\nano\second\per\byte}$ differ by 4\,\%, while $\Delta = \SI{0.0130}{\nano\second\per\byte}$ is 28\,\% of a clean block's cost. The Loongson-3A6000 is similar: the two clean-block costs agree to within 6\,\%, and $\Delta = \SI{0.0404}{\nano\second\per\byte}$.
That is why our ARM~NEON and LSX kernels have no direct fast path and always write blocks to the stack buffer. NEON's decoding step uses the structured load \texttt{vld4}, which de-interleaves four byte streams from memory, whereas classification wants four contiguous 16-byte loads. Decoding a block straight from the registers that hold it would require an in-register transpose; on these cores, storing 64~bytes to an L1-resident buffer and reloading them with \texttt{vld4} is cheaper.

\paragraph{Short Lines.}
Equation~\ref{eq:profile} does not cover $L < 64$. There $f$ is capped at~1, but blocks hold two or more line breaks (about a third of them at $L=48$, nearly all at $L \leq 32$), so the single-whitespace path of Section~\ref{sec:blockloop} no longer applies and these blocks take the general table-driven compaction. On the Apple~M4~Max the cost keeps rising past $f=1$: \SI{0.061}{\nano\second\per\byte} at $L=64$, \SI{0.071}{} at $L=48$, \SI{0.075}{} at $L=32$ and \SI{0.086}{\nano\second\per\byte} at $L=16$. On AVX-512, where \texttt{vpcompressb} costs the same for any mask, it stays between \SI{0.048}{} and \SI{0.053}{\nano\second\per\byte}.

\paragraph{Instruction-Level Parallelism.}
Dividing instructions per byte by cycles per byte in Table~\ref{tab:perf-counters} gives the instructions retired per cycle. Our NEON kernel retires 5.3--5.8 instructions per cycle on the Apple~M4~Max on bing, googlelogo and email, and our AVX-512 kernel 5.2--5.8 on the AMD~EPYC~9R45 (Zen~5) on bing and googlelogo: the 64-character block exposes enough independent work---four 16-byte registers on NEON, and classification, translation and packing that do not depend on the previous block---to keep these wide cores busy. On the Intel Xeon~Gold~6548N the retirement rate is lower. These cores issue 512-bit vector operations on only two ports, 0 and~5, and byte shuffles and permutations (\texttt{vpermb}, \texttt{vpermi2b}, \texttt{vpcompressb}) on port~5 alone. On whitespace-free input a 64-character block takes 4.4~cycles and 10~instructions, and port~5 receives 4.5~micro-operations per block (the counter also includes a scalar port): it is busy essentially every cycle, while port~0 is busy 79\,\% of the time. Front-end and memory stalls are negligible (0.4\,\% and 1.2\,\% of issue slots).
At MIME line length ($L=76$) a block takes 11.8~cycles and 32~instructions. The extra work---compaction into the stack buffer, the reload and the decoding from it---lands on the same two vector ports, busy 84\,\% (port~5) and 69\,\% (port~0) of the cycles, and memory-bound slots stay at 6\,\%: the stack buffer stays in L1 and its round trip costs vector instructions, not memory latency. On NEON, by contrast, the whitespace penalty is the table-driven compaction ($\Delta$).

\begin{table}
  \caption{Timing profile in nanoseconds per input byte. The quantity $t_{\mathrm{fast}}$ is the cost of a clean block on the direct fast path, $t_{\mathrm{buf}}$ its cost in buffered mode, and $\Delta$ the extra cost of a block carrying an ignorable character. The cost $t(f{=}0.83)$ is measured at MIME line length. The last column is $f\Delta/t$, the share of decoding time attributable to compaction at that line length.}%
  \label{tab:profile}
  \centering
  \small
  \begin{minipage}{\textwidth}
    \centering
    \begin{tabular}{llrrrrr}
\toprule
\multirow{2}{*}{System} & \multirow{2}{*}{Kernel} & $t_{\mathrm{fast}}$ & $t_{\mathrm{buf}}$ & $\Delta$ & $t(f{=}0.83)$ & whitespace \\
 & & (ns/B) & (ns/B) & (ns/B) & (ns/B) & share at MIME \\
\midrule
Apple M4 Max & NEON (128-bit) & 0.0464 & 0.0483 & 0.0130 & 0.0591 & 18\,\% \\
Intel Xeon Gold 6548N & AVX-512 VBMI2 (512-bit) & 0.0199 & 0.0448 & 0.0034 & 0.0469 & 6\,\% \\
Loongson-3A6000 & LASX (256-bit) & 0.1569 & 0.1667 & 0.0404 & 0.1966 & 17\,\% \\
\bottomrule
\end{tabular}
  \end{minipage}
\end{table}

\subsection{C++ Ablation}
\label{sec:ablation}

We ablate the two design parameters of Section~\ref{sec:blockloop}.

\paragraph{The Single-Whitespace Path.}
When the classification mask has exactly one set bit, we build the shuffle indices arithmetically rather than loading them from a table (Section~\ref{sec:x64impl}). Removing this specialization, so that every block with whitespace takes the general table-driven path,\footnote{\masked{The modified code is available online at \url{https://github.com/simdutf/simdutf/tree/ablation}.}{Repository URL omitted for anonymous review.}} costs nearly 30\,\% on the Apple~M4~Max: in that rebuild the email dataset falls from \SI{15.2}{\giga\byte\per\second} to \SI{10.7}{\giga\byte\per\second}.
The size of the effect is expected from Section~\ref{sec:timingprofile}: with 76-character lines, 83\,\% of blocks contain an ignorable character and almost all of those contain exactly one, so the specialized path handles most of the blocks that compact at all.

\paragraph{Stack Buffer Size.}
The buffer must hold at least 128~bytes; beyond that, the size is a parameter. Fig.~\ref{fig:bufsweep} sweeps it from 128 to \SI{2048}{\byte} at MIME line length on two systems: the Apple~M4~Max runs the NEON kernel with \emph{128-bit} registers, the Intel Xeon~Gold~6548N the AVX-512 kernel with \emph{512-bit} registers. Both flatten around \SI{512}{\byte}.

The mechanism is amortization. With a buffer of $N \times 64$~bytes, each flush decodes $N-1$ groups and performs one copy of the fewer than 64~bytes left over, so the bookkeeping cost per decoded group falls like $1/(N-1)$.  How much it matters differs: at \SI{128}{\byte} ($N=2$) the AVX-512 kernel drops from \SI{19.2}{\giga\byte\per\second} to \SI{5.9}{\giga\byte\per\second}, a factor of $3.2$, while the NEON kernel drops only from \SI{16.9}{\giga\byte\per\second} to \SI{15.3}{\giga\byte\per\second}. That matches the timing profile: on AVX-512 the buffer traffic dominates the whitespace penalty, on NEON the compaction does.
We use $N = 6$ (\SI{384}{\byte}) in the AVX-512 and LASX kernels and $N = 10$ (\SI{640}{\byte}) in the NEON and LSX kernels. Beyond \SI{512}{\byte} the speed is insensitive to the exact size; a larger buffer only wastes stack and cache.

\subsection{C++ Scalability Results}
\label{sec:scalability}

\begin{figure}
    \centering
    \includegraphics[width=0.55\linewidth]{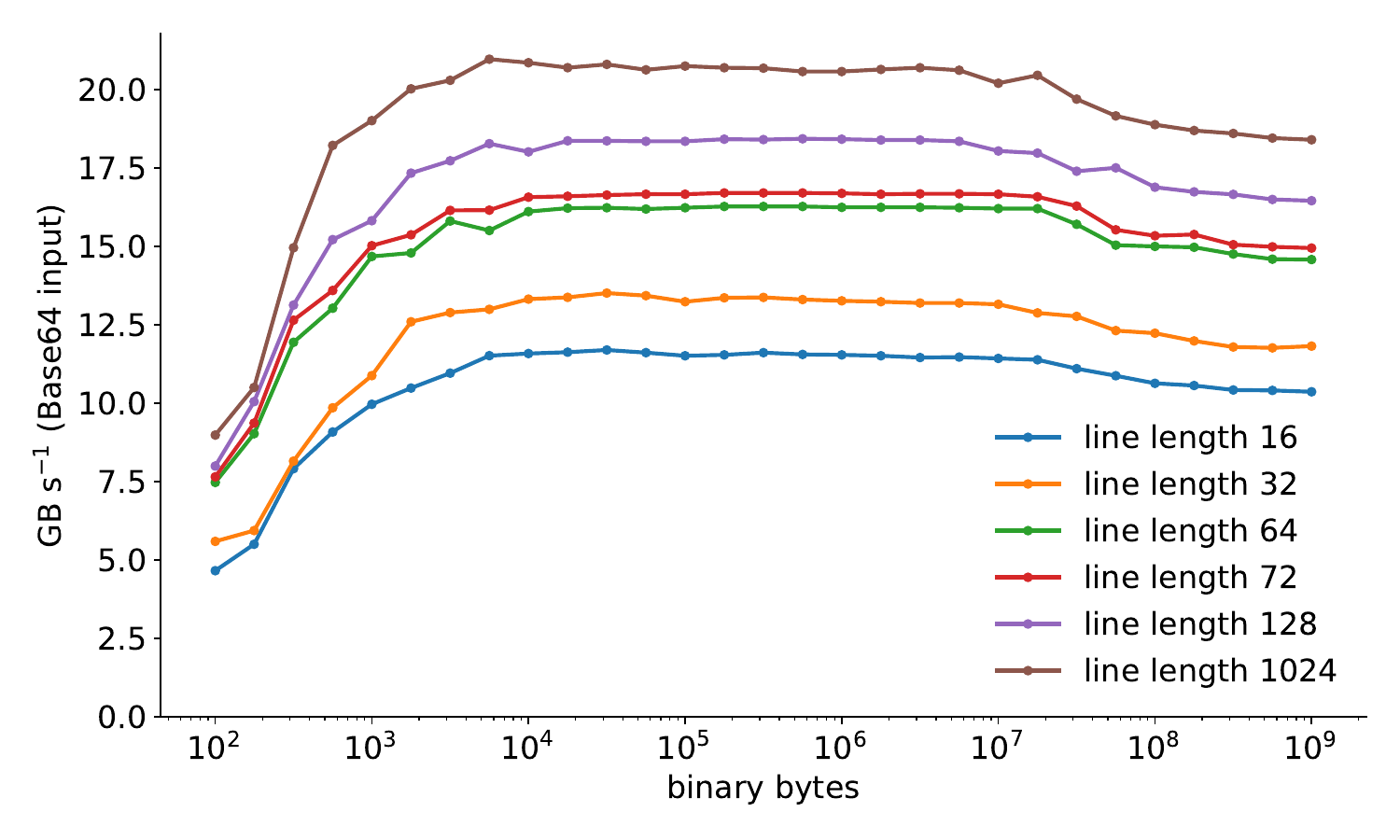}
    \caption{Decoding speed on the Apple~M4~Max for synthetic inputs of a fixed number of binary bytes (x-axis, logarithmic), Base64-encoded with a newline every 16, 32, 64, 72, 128 or 1024~characters. Speed is measured per Base64 input byte, including the newlines, using the fastest of repeated runs.}
    \label{fig:scalability_plot}
\end{figure}

To see how performance varies with input size and with the density of ignorable characters, we Base64-encoded random binary inputs from 1~kilobyte to 1~gigabyte---yielding encoded inputs of 1.3~kilobytes to 1.3~gigabytes---inserting a newline every 16, 32, 64, 72, 128 or 1024~characters.\footnote{\masked{The benchmark is available online at \url{https://github.com/lemire/base64scale}}{Benchmark URL omitted for anonymous review.}}
Fig.~\ref{fig:scalability_plot} gives the results on the Apple~M4~Max. Decoding speed rises with input size until a few kilobytes and is then stable until hundreds of megabytes, where a moderate slowdown appears as the data outgrows the \SI{32}{\mega\byte} L2 cache shared by the high-performance cores.
In the stable region the speeds match the timing profile of Section~\ref{sec:timingprofile}. As Equation~\ref{eq:profile} predicts, 72-character lines are about 20\,\% slower than 1024-character lines. Sixteen-character lines are 1.8 times slower than 1024-character lines, because most blocks then hold several line breaks, a case the equation does not cover.

\subsection{Node.js Results}
\label{sec:nodejs}

Major JavaScript runtimes rely on our implementation for ES2026 Base64 decoding. We wrote a JavaScript benchmark for Node.js,\footnote{\masked{\url{https://github.com/lemire/fromBase64}}{Benchmark URL omitted for anonymous review.}} which calls the v8 engine used by Chromium browsers; a similar browser benchmark is available online.\footnote{\masked{\url{https://simdutf.github.io/browserbase64/}}{Browser-benchmark URL omitted for anonymous review.}}
It decodes with the native \texttt{Uint8Array.fromBase64}, falling back on \path{new Uint8Array(Buffer.from(str,'base64'))} for older Node.js versions. To limit garbage-collector jitter we pass the flags \path{--max-old-space-size=1000}, \path{--max-semi-space-size=512} and \path{--noconcurrent_sweeping}, and after loading the data we measure 1000~iterations following a 1000-iteration warm-up, reporting the median.
\begin{table}
    \centering
    \caption{JavaScript decoding speed in gigabytes per second (Enron dataset, Node.js~25.5 (ours) vs. Node.js~19.9 (conventional))}
\begin{tabular}{lccc}
\toprule
                  & Node.js~19.9  & Node.js~25.5 (ours)  & speed-up\\\midrule
 Intel Xeon Gold 6548N    &       \SI{1.2}{\giga\byte\per\second}                 &     \SI{7.8}{\giga\byte\per\second}     & $6 \times$   \\
Apple M4 Max          &      \SI{2.0}{\giga\byte\per\second}                  &     \SI{9.3}{\giga\byte\per\second}  & $4.7 \times$       
       \\ \bottomrule
\end{tabular}
    \label{tab:nodejs}
\end{table}

We show our results in Table~\ref{tab:nodejs}. These numbers reflect decoding speeds using the Enron email dataset. On the Intel Xeon Gold 6548N system, decoding throughput increases by over \SI{6}{\times} (from \SI{1.2}{\giga\byte\per\second} in Node.js~19.9 to \SI{7.8}{\giga\byte\per\second} in Node.js~25.5), while on the Apple M4 Max, we observe a \SI{4.7}{\times} gain (from \SI{2.0}{\giga\byte\per\second} to \SI{9.3}{\giga\byte\per\second}). These gains stem primarily from the adoption of our high-performance Base64 decoding routines, which Node.js~25+ uses through the native \texttt{Uint8Array.fromBase64} method. Older Node.js versions, such as Node.js~19, fall back to slower scalar implementations.

Decoding in Node.js~25 remains slower than the raw C++ speed on the same hardware and data: \SI{7.8}{\giga\byte\per\second} versus \SI{18.4}{\giga\byte\per\second} on Intel, and \SI{9.3}{\giga\byte\per\second} versus \SI{14.7}{\giga\byte\per\second} on Apple, or 37--58\,\% slower. Part of this is unavoidable, since the JavaScript function allocates a new \texttt{Uint8Array} for every call whereas our C++ benchmark decodes into a pre-allocated buffer. Our profiling attributes the rest to data copying and initialization: in the v8 engine\footnote{\url{https://github.com/v8/v8/}} the decoded data is first written to a newly allocated \path{std::unique_ptr<char[]>} and then copied into a \path{MaybeDirectHandle<JSArrayBuffer>}. We expect future v8 releases to remove that copy.

\subsection{In-Browser Benchmarks}
\label{sec:browser}

Browsers expose the same \texttt{Uint8Array.fromBase64} API\@. On the Apple~M4~Max we benchmark two production engines: Firefox fully headless, and Safari through \texttt{safaridriver}. Safari's WebKit uses our decoder, whereas Firefox's SpiderMonkey has an independent implementation and serves as a reference. On the Enron email data Safari reaches \SI{10.2}{\giga\byte\per\second} against Firefox's \SI{3.3}{\giga\byte\per\second}: below the native C++ figures, but the engine that adopts our decoder is $3\times$ faster than the independent one. Chromium (V8), which also ships our decoder, reaches \SI{6.6}{\giga\byte\per\second} on the same machine and data.


\section{Conclusion}
\label{sec:conclusion}

We have described a SIMD Base64 decoder that implements the ES2026/\textsc{WHATWG} forgiving-base64 rules of Section~\ref{sec:conformance}---alphabet selection, ignorable whitespace, padding placement, and \texttt{lastChunkHandling}. Prior SIMD decoders assumed whitespace-free input. Against Turbo-Base64 preceded by our own vectorized whitespace-removal pass, we are up to $3.5\times$ faster on clean input (\SI{97.6}{\giga\byte\per\second} against \SI{28.0}{\giga\byte\per\second} on Zen~5) and up to $2.5\times$ on whitespace-containing input (\SI{5.2}{\giga\byte\per\second} against \SI{2.1}{\giga\byte\per\second} on the Loongson-3A6000, Tables~\ref{tab:competitors}--\ref{tab:competitors-aws-x64}). Against \texttt{base64-simd}, the only other library with forgiving semantics, we are 2.2--14 times faster on whitespace-containing input. 

The cost of the forgiving rules is modest and depends on the architecture (Section~\ref{sec:timingprofile}). At MIME line lengths, whitespace handling accounts for 18\,\% of decoding time on ARM~NEON and 17\,\% on LoongArch~LASX, where the compaction dominates. On AVX-512~VBMI2 the compaction is essentially free---\texttt{vpcompressb} is one instruction---and almost the whole penalty is the loss of the direct-to-output fast path once the stack buffer becomes non-empty. That is why our NEON and LSX kernels always buffer while the AVX-512, AVX2 and LASX kernels do not. On wide-vector machines, the store-to-load round trip through the stack buffer is the remaining bottleneck, not the compaction.

Since mid-2025 our SIMD decoder has shipped in Safari, Chromium-based browsers, and major runtimes (Node.js~25+, Deno, and Bun), backing the ES2026 \texttt{Uint8Array.fromBase64} and \texttt{Uint8Array.toBase64} APIs.

The natural next targets are the scalable vector extensions discussed in Section~\ref{sec:scalable}: RVV, where \texttt{vcompress.vm} supplies byte-granular compaction and the segment loads match NEON's structured loads, and SVE2, which would simplify predication and tail handling but has no byte-granular \texttt{COMPACT} until SVE2.2.

\ifanon\else
\begin{acks}
We received funding from the Natural Sciences and Engineering Research
Council of Canada, Grant Number:
RGPIN-2024-03787.
Our work would not have been possible
without the simdutf library community, including Wojciech Muła, Yagiz Nizipli, and Paul Dreik.
Muła provided an optimization for the case
where whitespace characters are isolated.
Nizipli optimized the Google v8 implementation.
Dreik provided crucial testing code.
Our work would also not have been possible
without the collaboration of the Apple WebKit and Google v8 engineers.
\end{acks}
\fi

\bibliographystyle{ACM-Reference-Format}
\bibliography{base64whatwgacm}

\end{document}